\documentclass[aps,showpacs,preprintnumbers,amsmath,amssymb]{revtex4}
\usepackage{dcolumn}
\usepackage[dvips]{epsfig}
\usepackage{graphicx}
\usepackage{amssymb}
\usepackage{color}
\usepackage{enumerate}
\usepackage{subfigure}
\newcommand{\degree}{^\circ}

\begin{document}
\baselineskip=0.8 cm
\title{Chaotic shadow of a non-Kerr rotating compact object with quadrupole mass moment}
\author{Mingzhi Wang$^{1}$,  Songbai Chen$^{1,2,3,4}$\footnote{Corresponding author: csb3752@hunnu.edu.cn}, Jiliang
Jing$^{1,2,3,4}$ \footnote{jljing@hunnu.edu.cn}
}
\affiliation{$ ^1$Institute of Physics and Department of Physics, Hunan
Normal University,  Changsha, Hunan 410081, People's Republic of
China \\ $ ^2$Key Laboratory of Low Dimensional Quantum Structures \\
and Quantum Control of Ministry of Education, Hunan Normal
University, Changsha, Hunan 410081, People's Republic of China\\
$ ^3$Synergetic Innovation Center for Quantum Effects and Applications,
Hunan Normal University, Changsha, Hunan 410081, People's Republic
of China\\
$ ^4$Center for Gravitation and Cosmology, College of Physical Science and Technology,
Yangzhou University, Yangzhou 225009, China}

\begin{abstract}
\baselineskip=0.6 cm
\begin{center}
{\bf Abstract}
\end{center}

We have studied numerically the shadows of a non-Kerr rotating compact object with quadrupole mass moment, which belongs to  Manko-Novikov family. The non-integrable photon motion caused by quadrupole mass moment affects sharply the shadow of the compact object. As the deviation parameter related to quadrupole mass moment is negative, the shadow of compact object is prolate and there are two disconnected main shadows with eyebrows located symmetrically on both sides of the equatorial plane. As the deviation parameter is positive, the shadow becomes oblate and the main shadow is joined together in the equatorial plane. Moreover, in this positive cases, there is a disorder region in the left of shadow which increases with the quadrupole-deviation parameter. Interestingly, we also find that  Einstein ring is broken as the deviation from Kerr metric is larger than a certain critical value. This critical value decreases with the rotation parameter of black hole. Especially,  the observer on the direction of rotation axis will find some concentric bright rings in the black disc.  Finally, supposing that the gravitational field of the supermassive central object of the galaxy described by this metric, we estimated the numerical values of the observables for the black hole shadow.

\end{abstract}
\maketitle

\newpage
\section{Introduction}

The current observations of gravitational waves \cite{gw,gw1,gw2,gw3,gw4} confirm  the existence of black holes in our Universe. According to the no-hair theorem \cite{noh}, one can find that a neutral rotating
black hole in asymptotically flat and matter-free spacetime is
described entirely by the Kerr metric, which is characterized uniquely
by a mass and an angular momentum, and then all
rotating astrophysical black holes in our Universe should be Kerr black
holes. However, there is no direct evidence to indicate that black hole candidates must be Kerr black holes, even if the current observations of gravitational waves \cite{gw,gw1,gw2,gw3,gw4} cannot exclude the possibility that the geometry of these candidates significantly deviates from the Kerr metric. Therefore, studying in detail the properties of a generic non-Kerr rotating black hole  and the corresponding observable effects become very significant \cite{TJo,RA2,RA20} since they can help us to understand features of black hole and to examine further the no-hair theorem.

One of important non-Kerr rotating asymptotically flat  spacetimes is Manko-Novikov spacetime \cite{MN}, which is an exact stationary and axial symmetric solution of the vacuum Einstein equations. Generally, besides the mass and rotation parameters, Manko-Novikov spacetime also possesses an infinite number of free parameters related to multipole moments, which describe the deviations away from the Kerr metric.
Here, we focus on only a particular subclass of the Manko-Novikov metric with a single deviation parameter denoting the difference from the Kerr case at the mass quadrupole order \cite{prd7}. Although there exist the naked singularity and the closed timelike curves, such a Manko-Novikov spacetime still is a good test-bed for exploring various observable consequences of deviations from the Kerr metric since the non-integrable particle motion originating from  the quadrupole-deviation
parameter does not appear in the Kerr black hole spacetime \cite{MN,prd7, an, non}, and the corresponding observable properties of non-integrable orbits  could help us to further test the Kerr black hole hypothesis.

Black hole shadow is a two-dimensional dark region in the observer's sky corresponding to light rays that fall into an event horizon when propagated backwards in time. It is well known that the shape and size of shadow carry the fingerprint of the geometry around the black hole \cite{sha1,sha2,sha3} and then
the shadow can be regarded as a potential tool to probe black hole parameters in the future observation, such as Event Horizon Telescope \cite{charge2,charge3} and European Black Hole Cam \cite{charge4}. The shadow of a Schwarzschild black hole is a perfect black disk, but for a rotating black hole, it becomes an elongated silhouette due to the so-called dragging effect \cite{sha2,sha3}. The cusp silhouette of shadow emerges in the spacetime of a Kerr black hole with Proca hair \cite{fpos2} and of a Konoplya-Zhidenko rotating non-Kerr black hole \cite{sb10} as the suitable spacetime parameters are selected. Especially, the self-similar
fractal structures are found in the black hole shadow for the cases in which the photon motion is not variable-separable, such as, a Kerr black hole with scalar hair \cite{sw,swo,astro,chaotic}, a binary black hole system \cite{binary, sha18} or Bonnor black diholes with magnetic dipole moment \cite{my}. The recent investigation indicate that these novel structure and patterns in shadows are determined actually by the non-planar bound photon orbits \cite{fpos2} and  the invariant phase space structures \cite{BI} for the photon motion in the background spacetimes. Moreover, the shadows of black holes characterizing by other parameters have been studied recently \cite{swo7,sha4,sha5,sha6,sha7,sha9,sha10,sha11,sha12,sha13,sha14,sha15,sha16,
sb1,sha17,sha19,sha19s1,sha19s2,shan1,shan1add}, which indicate that these  parameters give rise to the richer silhouettes for black hole shadows in various theories of gravity.

In this paper, we will study the shadow of a non-Kerr rotating compact object with quadrupole mass moment \cite{MN, an, prd7}, which is Manko-Novikov spacetime with a single quadrupole-deviation parameter. Since the presence of the deviation parameter could lead to the chaotic motion of photon, it is expected that the pattern and structures of the shadow will have some essential features differed from that of Kerr black hole. The main purpose is to probe what new features of the shadow casted by the non-Kerr rotating compact object with quadrupole mass moment \cite{MN, an, prd7} and to detect the effect of quadrupole-deviation parameter on the shadow.

The paper is organized as follows. In Sec. II, we review briefly the Manko-Novikov metric and then analyze equation of photon motion in this spacetime. In Sec. III, with the backward ray-tracing method,  we present numerically the shadows for the non-Kerr rotating compact object with quadrupole mass moment and  probe the effect of quadrupole-deviation parameter on the shadow for the compact object. In Sec.IV, we suppose that the gravitational field of
the supermassive black hole at the center of our galaxy can
be described by the spacetime of a non-Kerr rotating black hole with quadrupole mass moment and then obtain the numerical
results for the observables about black hole shadow. Finally, we present a summary.

\section{Manko-Novikov Spacetime and null geodesics}

Manko-Novikov metric is an exact stationary, axial symmetric solution of the vacuum Einstein equations with extra higher-order moments which describe the deviations away from the Kerr metric \cite{MN, an, prd7}. Generally, Manko-Novikov solution has an infinite number of free parameters and the presence of these parameters could destroy
the event horizon which implies that it is not a black-hole spacetime within a certain range of parameters. Here, we focus on only a particular subclass of the Manko-Novikov metric depends on three parameters: the mass $M$, the spin $S $ and the dimensionless parameter $q$. The first two parameters are identical to those of the corresponding Kerr metric, but the third parameter $q$ measures the deviation of
Manko-Novikov quadrupole mass moment $Q$ from the Kerr quadrupole moment $Q_{Kerr}=-S^2/M$, i.e., $q=-(Q-Q_{Kerr})/M^3$. In the prolate spheroidal coordinates,  this Manko-Novikov metric can be described by a Weyl-Papapetrou line element as \cite{MN, an,non, prd7,4}:
\begin{eqnarray}
ds^2=-f(dt-\omega d\phi)^2+k ^{2}f^{-1}e^{2\gamma}(x^{2}-y^{2})\bigg(\frac{dx^{2}}{x^{2}-1}+\frac{dy^{2}}{1-y^{2}}\bigg)+k^{2}f^{-1}(x^{2}-1)(1-y^{2})d\phi^{2}
\label{MNdg}
\end{eqnarray}
where
\begin{eqnarray}
\alpha&=&\frac{-M+\sqrt{M^2-(S/M)^2}}{(S/M)}, \label{a}\\
k&=&M\frac{1-\alpha^2}{1+\alpha^2}, \label{k}\\
\beta&=&q \frac{M^{3}}{k^{3}}, \label{b}\\
f &=& e^{2 \psi}\frac{A}{B}, \label{ffunc} \\
\omega &=& 2 k e^{-2 \psi}\frac{C}{A}-4 k \frac{\alpha}{1-\alpha^2}, \\
e^{2 \gamma} &=& e^{2 \gamma^\prime}\frac{A}{(x^2-1)(1-\alpha^2)^2}, \label{fexpgam} \\
A &=& (x^2-1)(1+a~b)^2-(1-y^2)(b-a)^2,\label{fA} \\
B &=& [(x+1)+(x-1)a~b]^2+[(1+y)a+(1-y)b]^2,\label{fB} \\
C &=& (x^2-1)(1+a~b)[(b-a)-y(a+b)] \nonumber \\
&&+ (1-y^2)(b-a)[(1+a~b)+x(1-a~b)],
\end{eqnarray}
\begin{eqnarray}
\psi &=& \beta \frac{P_2}{R^3}, \label{fC}\\
\gamma^\prime &=& \ln{\sqrt{\frac{x^2-1}{x^2-y^2}}}+\frac{3\beta^2}{2 R^6}
(P_3^2-P_2^2) \nonumber \\ &+& \beta \left(-2+\displaystyle{\sum_{\ell=0}^2}
\frac{x-y+(-1)^{2-\ell}(x+y)}{R^{\ell+1}}P_\ell\right), \label{fgampr}\\
a &=& -\alpha \exp {\left[-2\beta\left(-1+\displaystyle{\sum_{\ell=0}^2}
\frac{(x-y)P_\ell}{R^{\ell+1}}\right)\right]}, \label{fa}\\
b &=& \alpha \exp {\left[2\beta\left(1+\displaystyle{\sum_{\ell=0}^2}
\frac{(-1)^{3-\ell}(x+y)P_\ell}{R^{\ell+1}}\right)\right]}, \label{fb}\\
R      &=& \sqrt{x^2+y^2-1}, \label{fR}\\
P_\ell &=& P_\ell (\frac{x~y}{R}), \label{fLegA}
\end{eqnarray}
and $P_\ell(z)$ are the $l$ order Legendre polynomials.  The parameter $q>0$ represents an oblate deviation of the Kerr metric, while $q < 0$ represents a prolate
deviation. As $q$ vanishes, the Manko-Novikov metric (\ref{MNdg}) reduces exactly to the Kerr metric. Another spacetime metric describing the gravitational field  of a compact object with quadrupole mass moment is the so-called quasi-Kerr metric \cite{mapping}, which is obtained by perturbing the Kerr metric with the help of the well-known Hartle-Thorne exterior metric \cite{map22}.  This quasi-Kerr metric is fully accurate up to the first order in terms of the dimensionless quadrupole moment parameter $q$ and the second order in terms of angular momentum. Although the quasi-Kerr metric  can be reduced to Kerr case as the parameter $q$ disappears, it is valid only on the slowly rotating cases. This is different from the Manko-Novikov metric in which there is no such a restriction on the rotation parameter.

With a coordinate change \cite{MN, an,non, prd7,4, 1, 2},
\begin{eqnarray}
x=\frac{r-M}{k},\;\;\;\;\;\;\;\;\;\;\;\;y=\cos\theta,
\label{bh}
\end{eqnarray}
the Manko-Novikov metric (\ref{MNdg}) can be rewritten as a form in the standard Boyer-Lindquist coordinates
\begin{eqnarray}
ds^2=-f(dt-\omega d\phi)^2+\frac{e^{2\gamma}\rho^{2}}{f\Delta}dr^{2}+\frac{e^{2\gamma}\rho^{2}}{f}d\theta^{2}+\frac{\Delta\sin^{2}\theta}{f}d\phi^{2},
\label{MNdg1}
\end{eqnarray}
where $\rho^{2}=(r-M)^{2}-k^{2}\cos^{2}\theta$, $\Delta=(r-M)^{2}-k^{2}$. And then the outer event horizon radius is $r_{h}=M+k$. The Hamiltonian of a photon motion along null geodesic in the spacetime (\ref{MNdg1}) can be expressed as
\begin{equation}
\label{hami}
\mathcal{H}(x,p)=\frac{1}{2}g^{\mu\nu}(x)p_{\mu}p_{\nu}
=\frac{1}{2}\bigg[\bigg(-\frac{1}{f}+
\frac{f\omega^{2}}{\Delta\sin^{2}\theta}
\bigg)p_{t}^{2}+\frac{f\Delta}{
e^{2\gamma}\rho^{2}}p_{r}^{2}
+\frac{f}{e^{2\gamma}\rho^{2}}p_{\theta}^{2}
+\frac{f}{\Delta\sin^{2}\theta}p_{\phi}^{2}
+2\frac{f\omega}{\Delta\sin^{2}\theta}p_{t}p_{\phi}\bigg]=0,
\end{equation}
where $p_{r}$ and $p_{\theta}$ are the components of momentum of the photon $p_{r}=g_{rr}\dot{r}$ and $p_{\theta}=g_{\theta\theta}\dot{\theta}$. Obviously, $t$ and $\phi$ are two cyclic coordinates for the spacetime (\ref{MNdg1}) since the metric functions are independent of these two coordinates. Therefore, there exist
 two conserved quantities $E$ and $L$ for the motion of photon, i.e.,
\begin{eqnarray}
\label{EL}
E=-p_{t}=f\dot{t}-f\omega\dot{\phi},\;\;\;\;\;\;\;\;\;\;\;\;\;\;
L=p_{\phi}=f\omega\dot{t}+\bigg(\frac{\Delta\sin^{2}\theta}{f}-f\omega^{2}\bigg)\dot{\phi},
\end{eqnarray}
which correspond to the energy  and the $z$-component of the angular momentum
of photon moving in the spacetime, respectively. With these two conserved quantities, we can find the null geodesic equations of photon can be expressed as
\begin{eqnarray}
\label{cdx}
\dot{t}&=&\frac{E}{f}+\frac{f\omega(L-E\omega)}{\Delta\sin^{2}\theta},\\
\ddot{r}&=&\frac{1}{2}\frac{\partial }{\partial r}\bigg[\ln\bigg(\frac{f\Delta}{e^{2\gamma}\rho^{2}}\bigg)\bigg]\dot{r}^{2}+\frac{\partial }{\partial \theta}\bigg[\ln\bigg(\frac{f}{e^{2\gamma}\rho^{2}}\bigg)\bigg]\dot{r}\dot{\theta}
+\frac{\Delta}{2}\frac{\partial }{\partial r}\bigg[\ln\bigg(\frac{e^{2\gamma}\rho^{2}}{f}\bigg)\bigg]\dot{\theta}^{2}\nonumber\\
&&-\frac{E^{2}\Delta f_{,r}}{2fe^{2\gamma}\rho^{2}}+\frac{f(L-E\omega)[2Ef\Delta \omega_{,r}+(L-E\omega)(f\Delta_{,r}-f_{,r}\Delta)]}{2e^{2\gamma}\rho^{2}\Delta\sin^{2}\theta}, \\
\ddot{\theta}&=&\frac{1}{2\Delta}\frac{\partial }{\partial \theta}\bigg[\ln\bigg(\frac{e^{2\gamma}\rho^{2}}{f}\bigg)\bigg]\dot{r}^{2}+\frac{\partial }{\partial r}\bigg[\ln\bigg(\frac{f}{e^{2\gamma}\rho^{2}}\bigg)\bigg]\dot{r}\dot{\theta}
+\frac{1}{2}\frac{\partial }{\partial \theta}\bigg[\ln\bigg(\frac{f}{e^{2\gamma}\rho^{2}}\bigg)\bigg]\dot{\theta}^{2}\nonumber\\
&&-\frac{f(L-E\omega)^{2}f_{,\theta}}{2e^{2\gamma}\rho^{2}\Delta\sin^{2}\theta}-\frac{E^{2}f_{,\theta}}{2fe^{2\gamma}\rho^{2}}+\frac{f(L-E\omega)[(L-E\omega)\cot\theta+E\omega_{,\theta}]}{e^{2\gamma}\rho^{2}\Delta\sin^{2}\theta},\\
\dot{\phi}&=&\frac{f(L-E\omega)}{\Delta\sin^{2}\theta},\label{cdx4}
\end{eqnarray}
with the constraint condition
\begin{equation}\label{emotion1}
H\equiv e^{2\gamma}\rho^{2}[\dot{r}^2+\Delta \dot{\theta}^2]-E^2\Delta+\frac{f^2}{\sin^{2}\theta}
\bigg(L-\omega E\bigg)^2=0,
\end{equation}
Obviously, the presence of quadrupole-deviation parameter $q$ yields that
the null geodesics equations are not be variable-separable because there is no existence of Carter-like constant in the Manko-Novikov spacetime (\ref{MNdg1}) and then only two integrals of motion $E$ and $L$ are admitted in this case. This implies that the motion of the photon could be chaotic, which should affect the shadow of a non-Kerr rotating compact object with quadrupole mass moment (\ref{MNdg1}).

\section{Shadow casted by a non-Kerr rotating compact object with quadrupole mass moment}

Let us now to study the shadows casted by a non-Kerr rotating compact object with quadrupole mass moment with "backward ray-tracing" method \cite{sw,swo,astro,chaotic,binary,sha18,my,BI,swo7}. In this method, the light rays are assumed to evolve from the observer backward in time and the information carried by each ray would be respectively assigned to a pixel in a final image in the observer's sky. With this spirit, we solve numerically the null geodesic equations (\ref{EL}) and (\ref{cdx}) for each pixel in the final image with the corresponding initial condition and obtain the image of shadow in observer's sky which is composed of the pixels connected to the light rays falling down into the horizon of black hole.
Considering that the spacetime of a non-Kerr rotating compact object with quadrupole mass moment(\ref{MNdg1}) is asymptotically flat, as in Refs.\cite{sw,swo,astro,chaotic,binary,sha18,my,BI,swo7}, one can expand the observer basis $\{e_{\hat{t}},e_{\hat{r}},e_{\hat{\theta}},e_{\hat{\phi}}\}$ as a form in the coordinate basis $\{ \partial_t,\partial_r,\partial_{\theta},\partial_{\phi} \}$
\begin{eqnarray}
\label{zbbh}
e_{\hat{\mu}}=e^{\nu}_{\hat{\mu}} \partial_{\nu},
\end{eqnarray}
where the transform matrix $e^{\nu}_{\hat{\mu}}$ obeys to $g_{\mu\nu}e^{\mu}_{\hat{\alpha}}e^{\nu}_{\hat{\beta}}
=\eta_{\hat{\alpha}\hat{\beta}}$, and $\eta_{\hat{\alpha}\hat{\beta}}$ is the metric of Minkowski spactime.
In general, the transformation (\ref{zbbh}) satisfied the spatial rotations and Lorentz boosts is not unique. For the Manko-Novikov  spacetime (\ref{MNdg1}), it is convenient to choice a decomposition associated with a reference
frame with zero axial angular momentum in relation to spatial infinity \cite{sw,swo,astro,chaotic,binary,sha18,my,BI,swo7,zero1}
\begin{eqnarray}
\label{zbbh1}
e^{\nu}_{\hat{\mu}}=\left(\begin{array}{cccc}
\zeta&0&0&\gamma\\
0&A^r&0&0\\
0&0&A^{\theta}&0\\
0&0&0&A^{\phi}
\end{array}\right),
\end{eqnarray}
where $\zeta$, $\gamma$, $A^r$, $A^{\theta}$, and $A^{\phi}$ are real coefficients.
From the Minkowski normalization
\begin{eqnarray}
e_{\hat{\mu}}e^{\hat{\nu}}=\delta_{\hat{\mu}}^{\hat{\nu}},
\end{eqnarray}
one can obtain
\begin{eqnarray}
\label{xs}
&&A^r=\frac{1}{\sqrt{g_{rr}}},\;\;\;\;\;\;\;\;\;\;\;\;\;\;\;\;
A^{\theta}=\frac{1}{\sqrt{g_{\theta\theta}}},\;\;\;\;\;\;\;\;\;\;\;\;\;\;\;
A^{\phi}=\frac{1}{\sqrt{g_{\phi\phi}}},\nonumber\\
&&\zeta=\sqrt{\frac{g_{\phi \phi}}{g_{t\phi}^{2}-g_{tt}g_{\phi \phi}}},\;\;\;\;\;\;\;\;\;\;\;\;\;\;\;\;\;\;\;\; \gamma=-\frac{g_{t\phi}}{g_{\phi\phi}}\sqrt{\frac{g_{\phi \phi}}{g_{t\phi}^{2}-g_{tt}g_{\phi \phi}}}.
\end{eqnarray}
Therefore, one can get the locally measured four-momentum $p^{\hat{\mu}}$ of a photon by the projection of its four-momentum $p^{\mu}$  onto $e_{\hat{\mu}}$,
\begin{eqnarray}
\label{dl}
p^{\hat{t}}=-p_{\hat{t}}=-e^{\nu}_{\hat{t}} p_{\nu},\;\;\;\;\;\;\;\;\;
\;\;\;\;\;\;\;\;\;\;\;p^{\hat{i}}=p_{\hat{i}}=e^{\nu}_{\hat{i}} p_{\nu}.
\end{eqnarray}
With the help of Eq.(\ref{xs}), the locally measured four-momentum $p^{\hat{\mu}}$ can be further written as
\begin{eqnarray}
\label{kmbh}
p^{\hat{t}}&=&\zeta E-\gamma L,\;\;\;\;\;\;\;\;\;\;\;\;\;\;\;\;\;\;\;\;p^{\hat{r}}=\frac{1}{\sqrt{g_{rr}}}p_{r} ,\nonumber\\
p^{\hat{\theta}}&=&\frac{1}{\sqrt{g_{\theta\theta}}}p_{\theta},
\;\;\;\;\;\;\;\;\;\;\;\;\;\;\;\;\;\;\;\;\;\;
p^{\hat{\phi}}=\frac{1}{\sqrt{g_{\phi\phi}}}L.
\end{eqnarray}
Repeating the similar operations in Refs.\cite{sw,swo,astro,chaotic,binary,sha18,my,BI,swo7},
one can obtain the position of photon's image in observer's sky
\begin{eqnarray}
\label{xd1}
x&=&-r_{obs}\frac{p^{\hat{\phi}}}{p^{\hat{r}}}
=-r_{obs}\frac{L}{\sqrt{e^{2\gamma}\rho^{2}
(\frac{\sin^{2}\theta}{f^{2}}-\frac{\omega^{2}}{\Delta})}\dot{r}}, \nonumber\\
y&=&r_{obs}\frac{p^{\hat{\theta}}}{p^{\hat{r}}}=
r_{obs}\frac{\sqrt{\Delta}\dot{\theta}}{\dot{r}},
\end{eqnarray}
where the spatial position of observer is set to ($r_{obs}, \theta_{obs}$).

In order to control the error in the numerical calculation of solving the coupled and complicated differential equations (\ref{cdx}-\ref{cdx4}), we adopt to the corrected fifth-order Runge-Kutta method \cite{Wu1,Wu2} in which the velocities $(\dot{r}, \dot{\theta})$ are corrected in integration and the numerical deviation is pulled back in a least-squares shortest path. This high precise method has been applied extensively to study chaotic motion  in various dynamical systems \cite{Wu1,Wu2,myte}. The energy of the dynamical system (\ref{cdx}) is subjected to the constraint $H=0$, and then $H$ could be regarded as a conserved quantity.  To ensure the high precision of the conserved quantity $H$ in the system of Eqs. (\ref{cdx})-(\ref{emotion1}) at every integration,
one can introduce a dimensionless parameter $\xi$ to make a connection between the numerical velocities $(\dot{r}, \dot{\theta})$ and the true value $(\dot{r}^{*}, \dot{\theta}^{*})$ in the form of
\begin{eqnarray}
\dot{r}^{*}=\xi\dot{r},\;\;\;\;\;\;\;\dot{\theta}^{*}=\xi\dot{\theta}.\label{scal1}
\end{eqnarray}
Inserting Eq.(\ref{scal1}) into Eq. (\ref{emotion1}), one can find that the scale factor of
velocity correction $\xi$ in the non-Kerr rotating compact object with quadrupole mass moment (\ref{MNdg1}) has a form
\begin{eqnarray}
\xi&=&\sqrt{\frac{E^2\Delta-\frac{f^2}{\sin^{2}\theta}
\bigg(L-\omega E\bigg)^2}{e^{2\gamma}\rho^{2}[\dot{r}^2+\Delta \dot{\theta}^2]}}.
\end{eqnarray}
\begin{figure}
\includegraphics[width=12cm ]{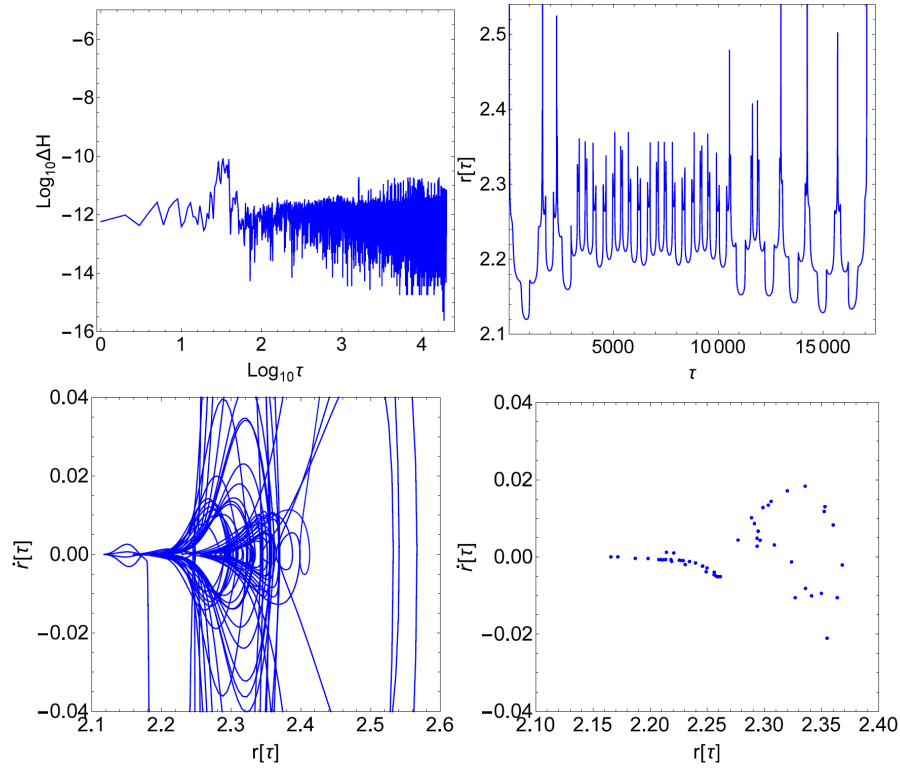}
\caption{ The error is controlled better than $10^{-10}$ by the velocity correction method (RK5+Correction) for a chaotic orbit of photon in the spacetime of a non-Kerr rotating compact object with quadrupole-deviation parameter $q=8$ and spin parameter $S=0.98$. The left and right panels in the top row correspond to the change of the error $H$ and the polar coordinate $r$ with time $\tau$, respectively. The panels in the bottom row  are the phase curve and Poincar\'{e} sections with $\theta=2.2$ on the plane $(r, \dot{r})$ for the corresponding chaotic orbit of photon, respectively. Here, we set the parameters $M=1$, $E=1$, $L=1.98341$,  and the initial conditions $\{r(0)=30; \dot{r}(0)=0.987395; \theta=\pi/2\}$.}
\label{s1}
\end{figure}
In this way, the conserved quantity $H$ in the system of Eqs. (\ref{cdx})-(\ref{emotion1}) can hold perfectly at every integration. Fig.\ref{s1} shows that the value of $H$ is
remained below $10^{-10}$ and then the error is controlled greatly, even for a chaotic orbit of photon, which
ensures that this method is very powerful so that it can avoid the pseudo chaos caused by numerical errors.
\begin{figure}
\includegraphics[width=16cm ]{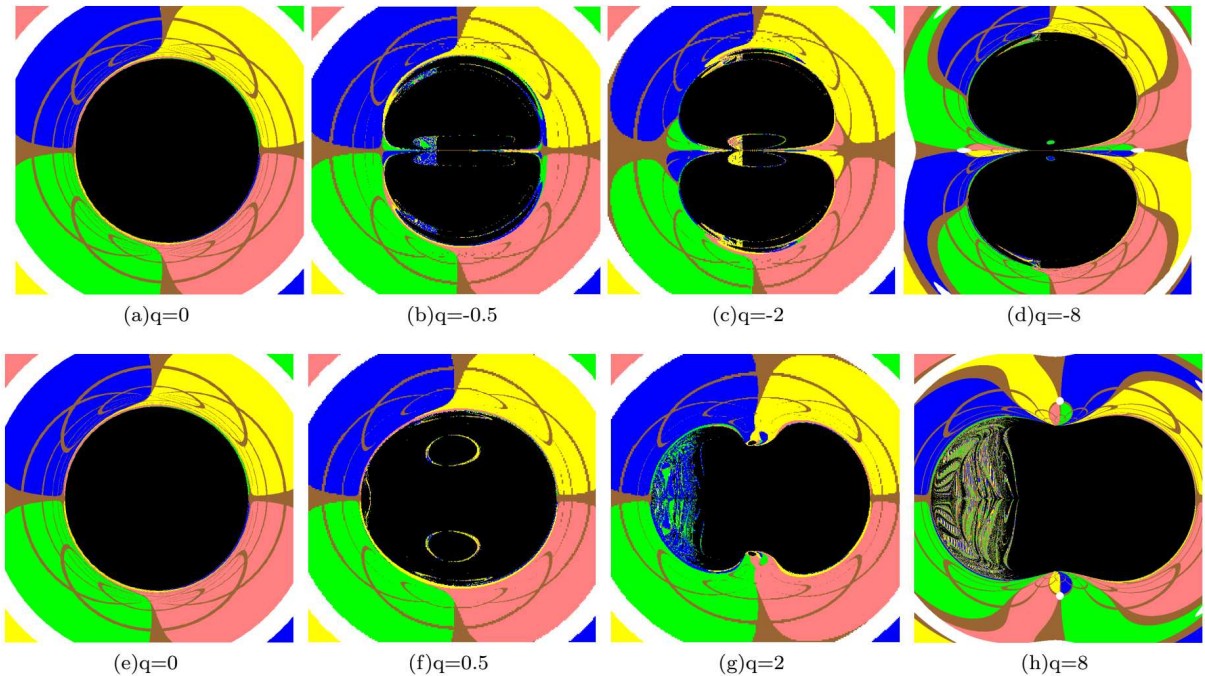}
\caption{The shadow of a non-Kerr rotating compact object with different quadrupole-deviation parameter $q$ for fixed spin parameter $S=0.2 M^{2}$. Here we set $M=1$ and the observer is set at $r_{obs}=30M$ with the inclination angle $\theta_{0}=90\degree$.}
\label{02}
\end{figure}

\begin{figure}
\includegraphics[width=16cm ]{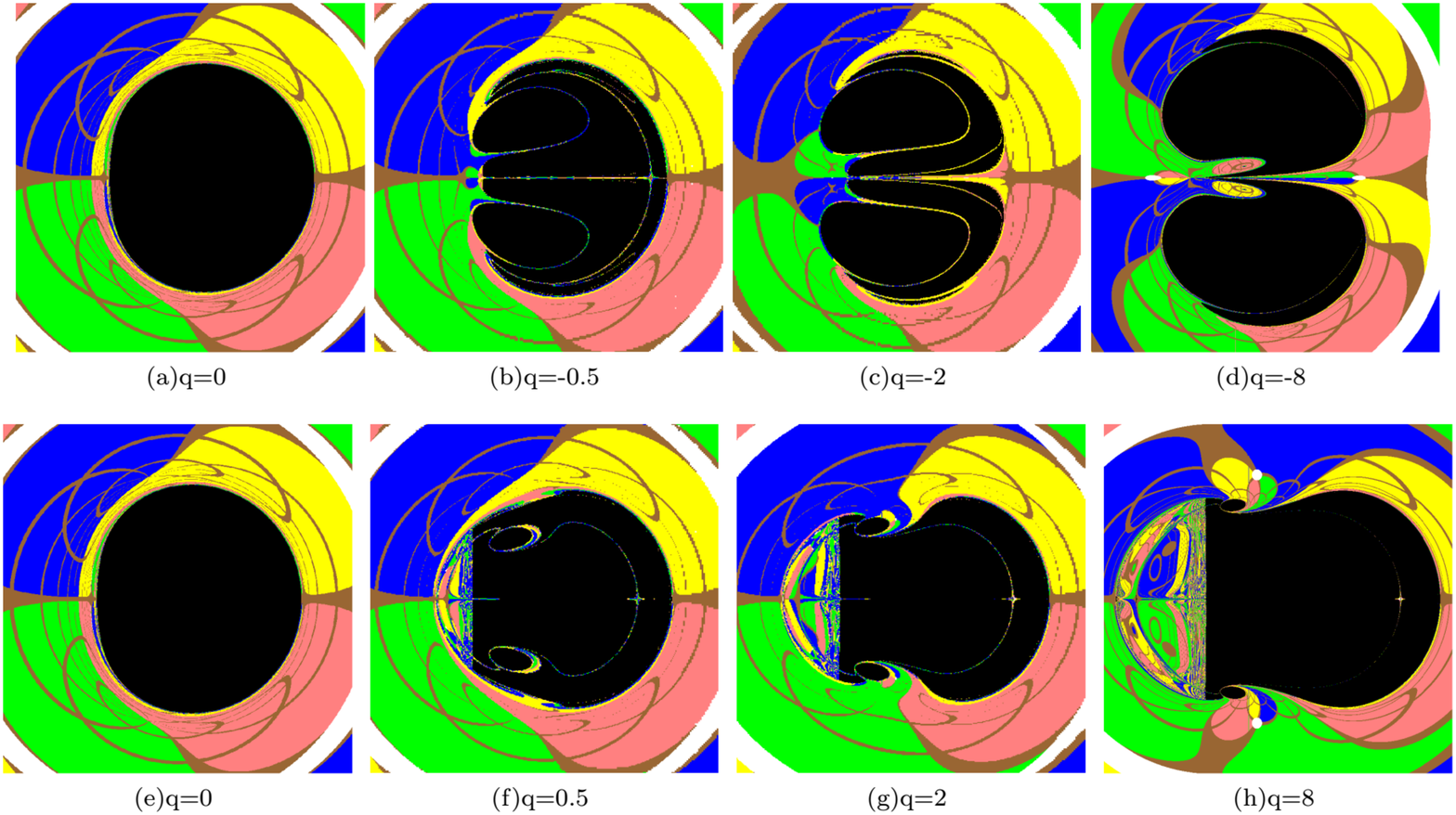}
\caption{The shadow of a non-Kerr rotating compact object with quadrupole-deviation parameter $q$ for fixed spin parameter $S=0.98 M^{2}$. Here we set $M=1$ and the observer is set at $r_{obs}=30M$ with the inclination angle $\theta_{0}=90\degree$.}
\label{098}
\end{figure}

In Figs.\ref{02}-\ref{098}, we present the shadow of a non-Kerr rotating compact object (\ref{MNdg1}) with different quadrupole-deviation parameter $q$  for the fixed spin parameter $S=0.2M^{2}$ and $S=0.98M^{2}$, respectively.
Here we set $M=1$ and the observer is set at $r_{obs}=30M$ with the inclination angle $\theta_{obs}=90\degree$.
From Fig.\ref{02}, we can find that the shadow of black hole is similar to a black disk in the case $q=0$ as discussed in Refs.\cite{sha2,sha3}. However, the presence of quadrupole-deviation parameter $q$ changes sharply the shape and pattern of the shadow of the non-Kerr rotating compact object.
Let us now firstly focus on the case with negative $q$. As $q=-0.5$, the shadow of the non-Kerr rotating compact object with quadrupole mass moment becomes prolate and it is split into two disconnected semicircular main shadows with eyebrows, which lie at symmetric positions above and below the equatorial plane. There exists an irregular bright region in each main shadow. Actually, we can  detect many other smaller shadows with a self-similar fractal structure, which is caused by chaotic lensing. As the deviation from Kerr black hole becomes larger, the curvature of the boundary line near the equatorial plane increases for the main shadow  and the irregular bright regions become smaller and move right. For the case with positive $q$, we find that the shadow of the non-Kerr rotating compact object with quadrupole mass moment becomes oblate and the main shadow are joined together in the equatorial plane.
As $q=0.5$, we find that two ring-like bright zones imbed symmetrically in the middle of the black shadow and an arc-like bright zone appears in the left of the shadow. Moreover, there exist the eyebrow-like shadows in the region near the north and south poles.   With the increase of the deviation, the shadow becomes oblate so that it turns into the shape of dumbbell for the larger $q$.
In addition, the left black region with the arc-like bright zone becomes a disorder region in which the black spots distribute dispersively and this disorder region increases with the quadrupole-deviation parameter $q$. These features of shadow means that the shape and pattern of the shadow of black hole with the positive $q$ are qualitatively different from those with negative $q$.
\begin{figure}
\includegraphics[width=16cm ]{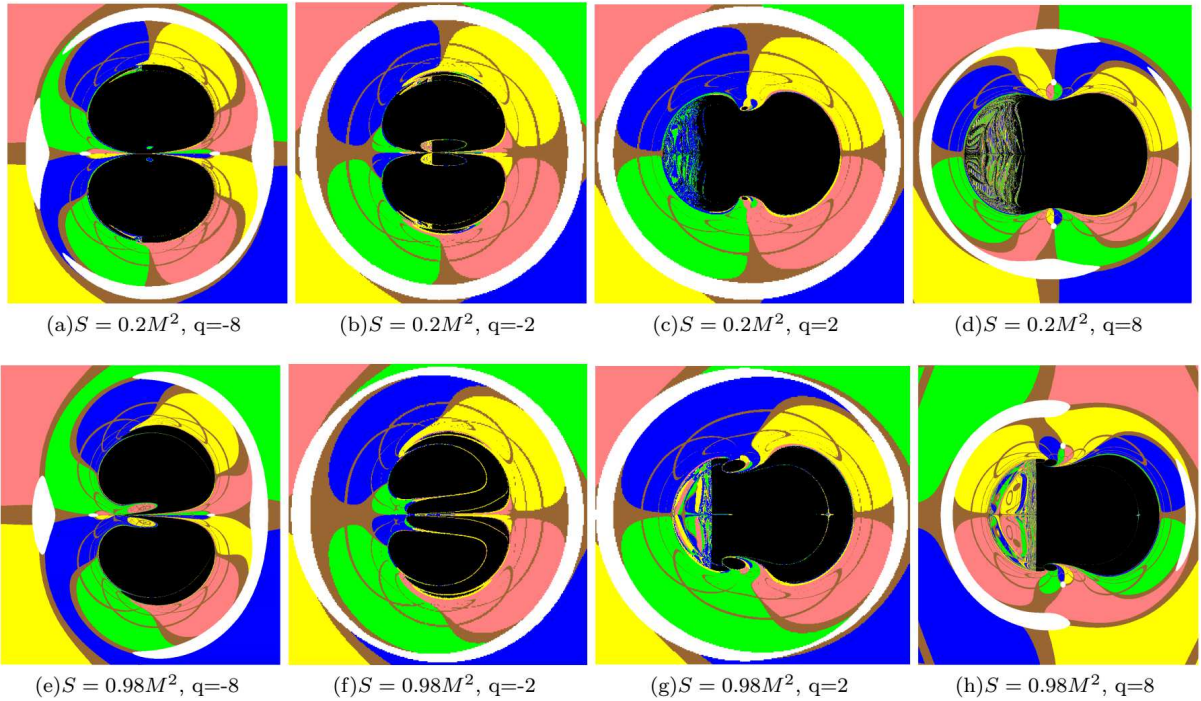}
\caption{The change of Einstein ring ( white ring in figures ) with the quadrupole-deviation parameter $q$ in the non-Kerr rotating black hole spacetime with quadrupole mass moment.}
\label{dd}
\end{figure}

In Fig.\ref{098}, we present the shadow of a rapidly rotating non-Kerr  compact object with quadrupole mass moment (\ref{MNdg1}). As $q=0$, the shadow has ``D"-type shape as expected since the metric (\ref{MNdg1}) is actually a rapidly rotating Kerr metric in this case. Comparing with the slowly rotating case, the shadow of a rapidly rotating non-Kerr  compact object with quadrupole mass moment has some new features. As $q=-0.5$, there are several bright curves which spilt the shadow into three main parts: two eyeball-like shadows, eyebrow-like shadows and intermediate black regions. With the increase of the deviation, we find that the two eyeball-like shadows increase, but the eyebrow-like shadows and intermediate black regions decrease so that the shadow becomes ``3"-type shape as $q=-8$. For the case with positive $q$, there also exist two eyeball-like shadows imbed in the main shadow with some smaller eyebrows.
\begin{table}[h]
\begin{center}
\begin{tabular}{|c|c|c|c|c|c|c|c|c|c|c|}
\hline \hline $S(M^{2})$ &$0.1 $ &\; 0.2 &\;0.3&\;0.4&\;0.5&\;0.6&\;0.7&\;0.8&\;0.9&\;0.98 \\
\hline
 $q_c(q>0)$ & 7.83& 7.25& 6.65&6.04& 5.44& 4.81&4.19&3.51&2.82&2.24
 \\\hline
 $q_c(q<0)$ & -7.83& -7.46& -7.05&-6.61& -6.14& -5.63&-5.06&-4.44&-3.72&-3.07\\
\hline\hline
\end{tabular}
\end{center}
\label{tab1} \caption{The critical value of quadrupole-deviation parameter $q_c$ accounting for a broken Einstein ring in the non-Kerr rotating black hole spacetime with quadrupole mass moment for different rotation parameter $S$.}
\end{table}
However, with the increase of quadrupole-deviation parameter $q$, the eyeball-like shadows become small and the main shadow first decreases and then increases. The disorder region in the left of shadow increases with the quadrupole-deviation parameter $q$, which is similar to that in the cases with the spin parameter $S=0.2M^{2}$.
In addition, from the subfigures (a), (d), (e) and (h) in Fig.\ref{dd}, we find that the Einstein ring ( white ring in figures ) is broken for the larger absolute value of $q$. The critical value of $q$ accounting for a broken Einstein ring is listed in Table I, which shows that the absolute value $|q_c|$  decreases with the rotation parameter of non-Kerr rotating black hole.
\begin{figure}
\includegraphics[width=16cm ]{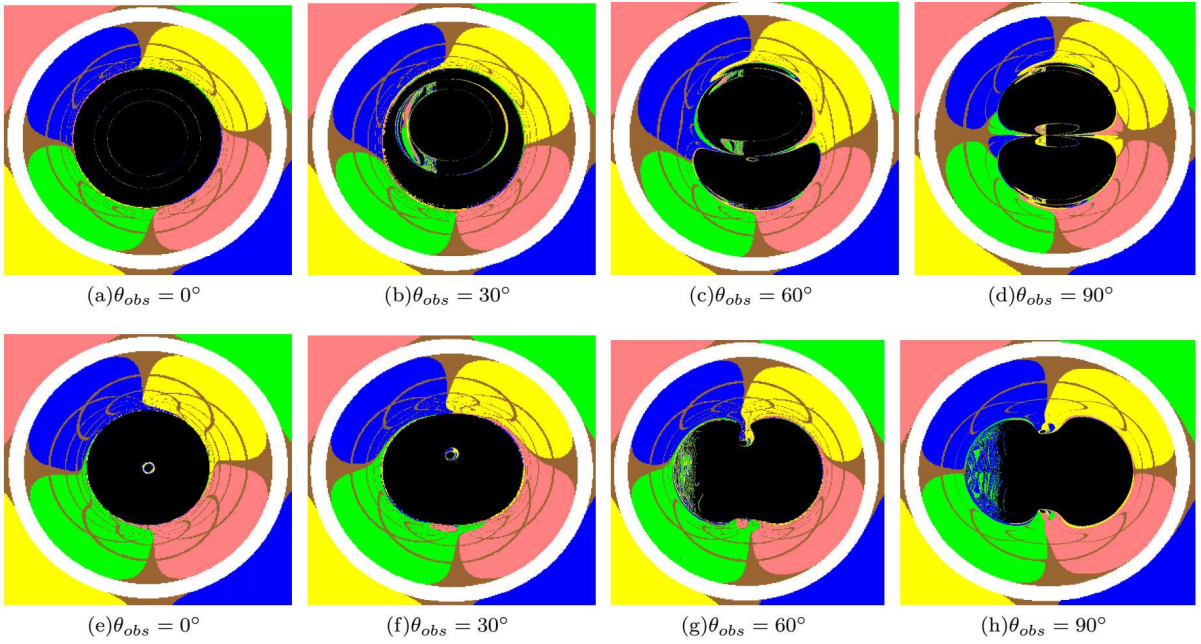}
\caption{The shadow of a non-Kerr rotating compact object with quadrupole mass moment for different observer's inclination angle $\theta_{obs}=0^{\circ}, 30^{\circ}, 60^{\circ}, 90^{\circ}$. The top row is for the quadrupole-deviation parameter $q=-2$ and the bottom row is for $q=2$. Here we set $M=1$ and $S=0.2M^{2}$.}
\label{022j}
\end{figure}
\begin{figure}
\includegraphics[width=16cm ]{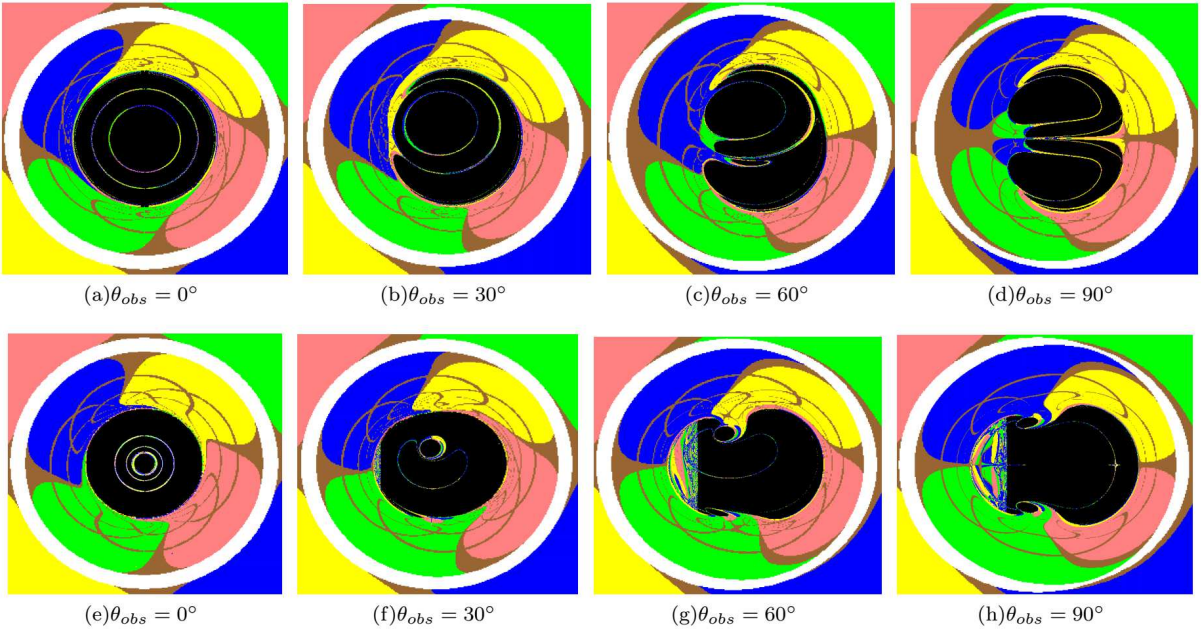}
\caption{The shadow of a non-Kerr rotating compact object with quadrupole mass moment for different observer's inclination angle $\theta_{obs}=0^{\circ}, 30^{\circ}, 60^{\circ}, 90^{\circ}$. The top row is for the quadrupole-deviation parameter $q=-2$ and the bottom row is for $q=2$. Here we set $M=1$ and $S=0.98M^{2}$.}
\label{0982j}
\end{figure}

In Figs. \ref{022j} and \ref{0982j}, we also present the dependence of the shadow on the observer's inclination angle $\theta_{obs}$ with the absolute value $|q|=2$ for $S=0.2M^{2}$ and $S=0.98M^{2}$, respectively. It is obvious that the shadow is center symmetric as $\theta_{obs}=0^{\circ}$. With the increase of the inclination angle $\theta_{obs}$, the center symmetry of shadow is gradually broken so that the shadows possess axial symmetric image as $\theta_{obs}=90^{\circ}$. Those properties are similar to those in
the Kerr black hole. However, we also find as $\theta_{obs}=0^{\circ}$,  there are concentric bright rings imbedded in the black disc, which is qualitatively different from that in the case of Kerr black hole. These distinct features in the shadow can be attributed to
the effect of quadrupole mass moment on the spacetime structures.

\section{Observational parameters for shadow of a non-Kerr rotating black hole with quadrupole mass moment }

In this section, supposing that the gravitational field of
the supermassive black hole at the galactic center of Milky
Way can be described by the non-Kerr rotating black hole with quadrupole mass moment, we estimate the numerical values of observables for the black hole shadow, and then we study the effect of the quadrupole-deviation parameter $q$ on these observables.

In order to characterize the shadow of a non-Kerr rotating black hole with quadrupole mass moment, we adopt six observables: the radius $R_{s}$, the oblateness parameter $K_{s}$, the three dimensionless distortion parameters ($\delta_{s,I}$, $\delta_{s,II}$, $\delta_{s,III}$) proposed in Ref.\cite{sha19s1}, and the ``thickness" parameter $T$ of shadow. The former two observables $R_{s}$ and $K_{s}$ can be
characterized by the points: the top position
($x_{t},y_{t}$), the bottom position ($x_{b},y_{b}$), the leftmost position ($x_{l},y_{l}$), and the rightmost position ($x_{r},y_{r}$) of the shadow.
The observable $R_{s}$ is defined as in \cite{sha9}
\begin{eqnarray}
\label{rs}
R_{s}=\frac{(x_{t}-x_{r})^{2}+y_{t}^{2}}{2|x_{t}-x_{r}|},
\end{eqnarray}
and the oblateness parameter $K_{s}$ of shadow can be defined as
\begin{eqnarray}
K_{s}\equiv\frac{\Delta x}{\Delta y}=\frac{x_{r}-x_{l}}{y_{t}-y_{b}}.
\end{eqnarray}
If a shadow is a standard circle, we find $K_{s}=1$. Moreover, the oblateness parameter satisfies $K_{s}>1$ for the case in which the contour of shadow is oblate and $K_{s}<1$ for the case where the shadow is prolate.

In order to describe further general characterization of shadow, A. Abdujabbarov \textit{et al} \cite{sha19s1} expanded the polar curve $R_{\psi}$ representing the shadow
as Legendre polynomials, i.e., $R_{\psi}=\sum\limits_{l=0}^{\infty}c_lP_l(\cos\psi)$, and defined three dimensionless distortion parameters
\begin{eqnarray}
&&\delta_{s,I}\equiv \frac{2\sum\limits_{l=1}^{\infty}c_{2l-1}}{\mathcal{B}},\nonumber\\
&&\delta_{s,II}\equiv \frac{2(\mathcal{B}^2-\mathcal{A}\mathcal{C})}{\mathcal{B}^2+\mathcal{A}^2},\nonumber\\
&&\delta_{s,III}\equiv 2\bigg(\sum_{l=0}^{\infty}c_lP_l(x_S)\bigg)\frac{\sum\limits_{l=0}^{\infty}c_lP_l(x_S)
-x_S\sum\limits_{l=1}^{\infty}c_{2l-1}-\mathcal{A}\mathcal{C}}{\mathcal{A}^2-
2x_S\sum\limits_{l=0}^{\infty}c_lP_l(x_S)+\bigg(\sum\limits_{l=0}^{\infty}c_lP_l(x_S)\bigg)^2},
\end{eqnarray}
where
\begin{eqnarray}
&&\mathcal{A}\equiv R_{\psi}(\psi=0)=\sum_{l=0}^{\infty}c_l,\nonumber\\
&&\mathcal{B}\equiv R_{\psi}(\psi=\pi/2)=\sum_{l=0}^{\infty}(-1)^l\frac{(2l)!}{2^{2l}(l!)^2}c_{2l},\nonumber\\
&&\mathcal{C}\equiv R_{\psi}(\psi=\pi)=\sum_{l=0}^{\infty}(-1)^lc_l.
\end{eqnarray}
and $x_S=\cos\psi_S$ which obeys the condition
\begin{eqnarray}
\frac{dR_{\psi}}{d\psi}\sin\psi+R_{\psi}\cos\psi=0.
\end{eqnarray}

For the aim to describe the multiple shadows appeared in Figs.1-5, we can scan across the shadow along certain a line  (such as image coordinate $x=const$ ) from the bottommost border of shadow to the equatorial plane and then define the
``thickness" parameter $T$ as
 \begin{eqnarray}\label{thick1}
T\equiv\sum_{i=1}^{n}l_i,
\end{eqnarray}
where $l_i$ is the width of $i$-th black shadow along the selected scan line and $n$ is total number of shadows.
If there is only a single shadow,  $T$ is a linear function of the image coordinate variable $y$ as we scan across the shadow along an arbitrary line. If there exist multiple shadows, one can find a scan line at least so that some platforms appear in the curve of $T (y)$, which can be called as ``plateau effect" (see in Figs.\ref{pt02}-\ref{pt098h3}). The fractal structures in these platforms distribution can be regarded as a potential observable effect of chaotic lensing.

\begin{figure}
\subfigure[$q=-2$]{ \includegraphics[width=4cm ]{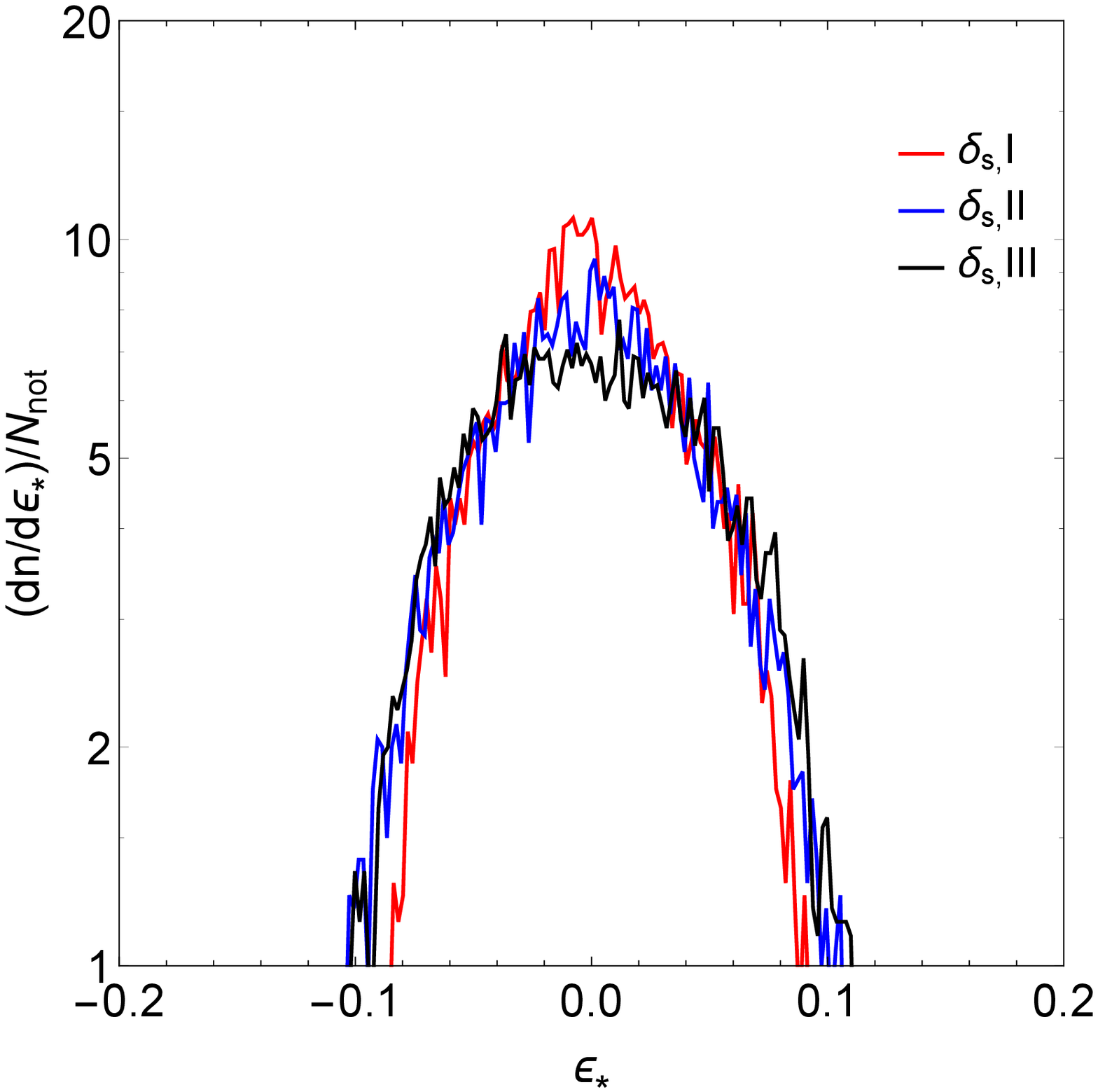}}\subfigure[$q=-0.5$]{ \includegraphics[width=4cm ]{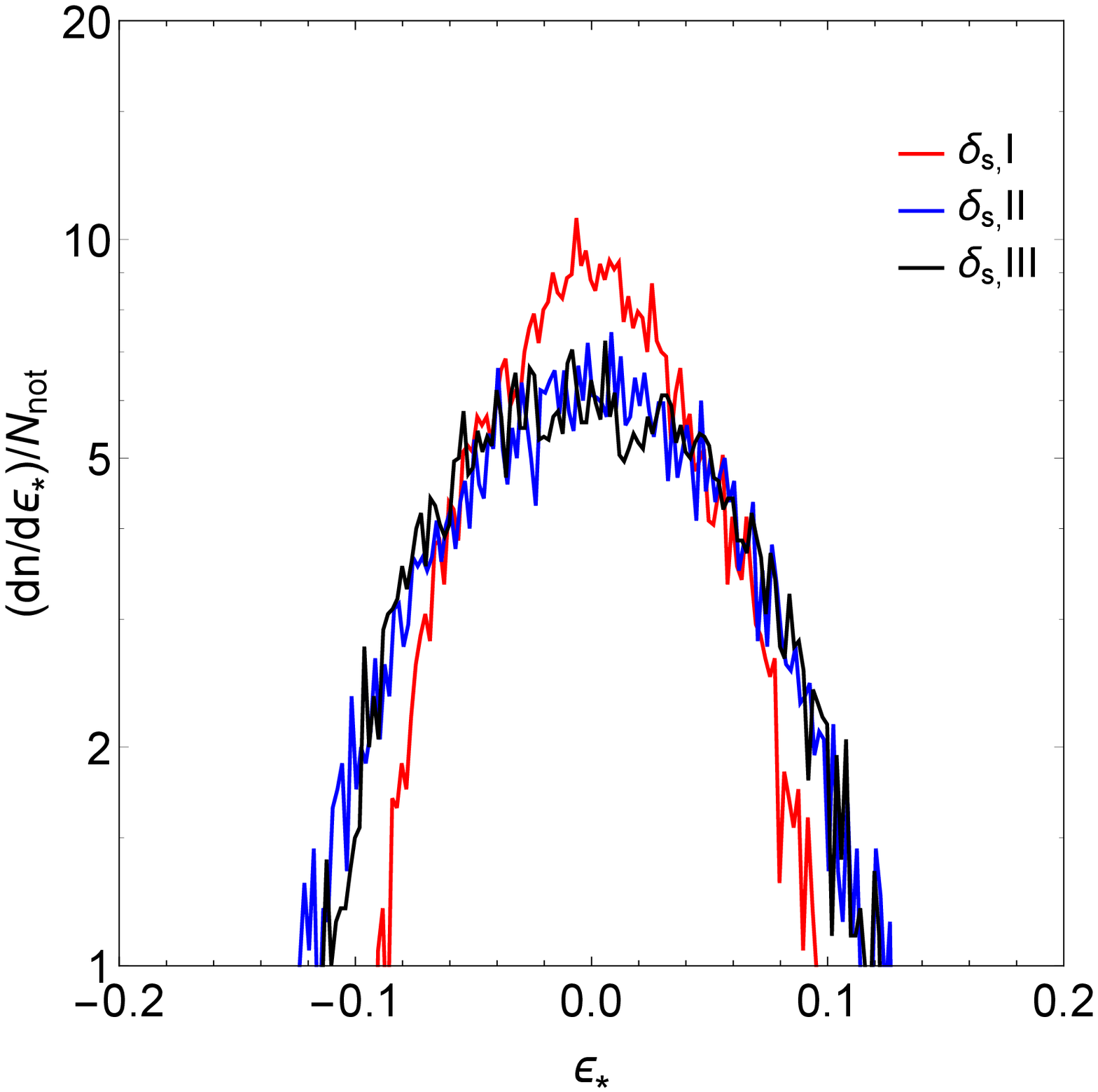}} \subfigure[$q=0.5$]{ \includegraphics[width=4cm ]{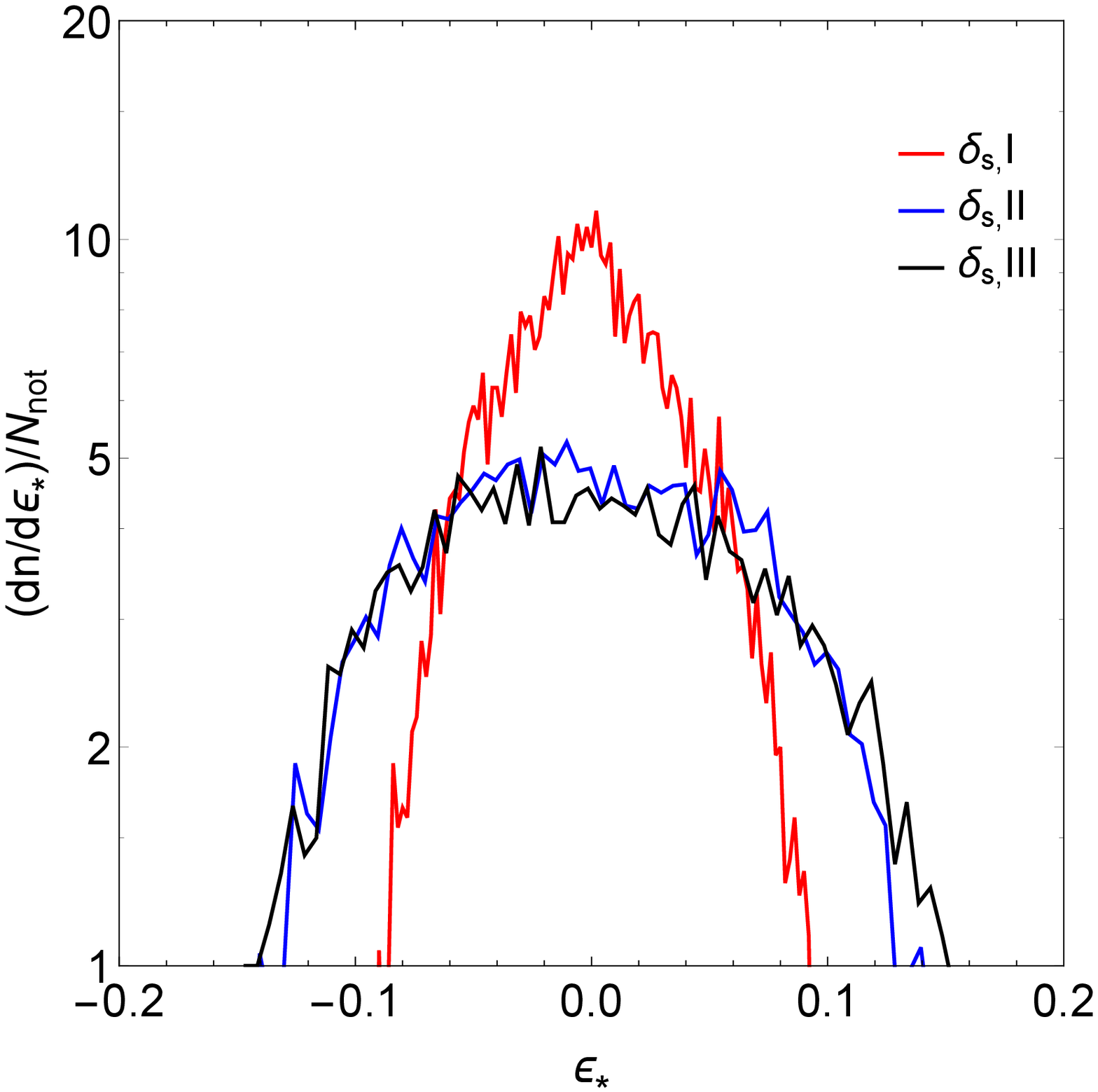}} \subfigure[$q=2$]{ \includegraphics[width=4cm ]{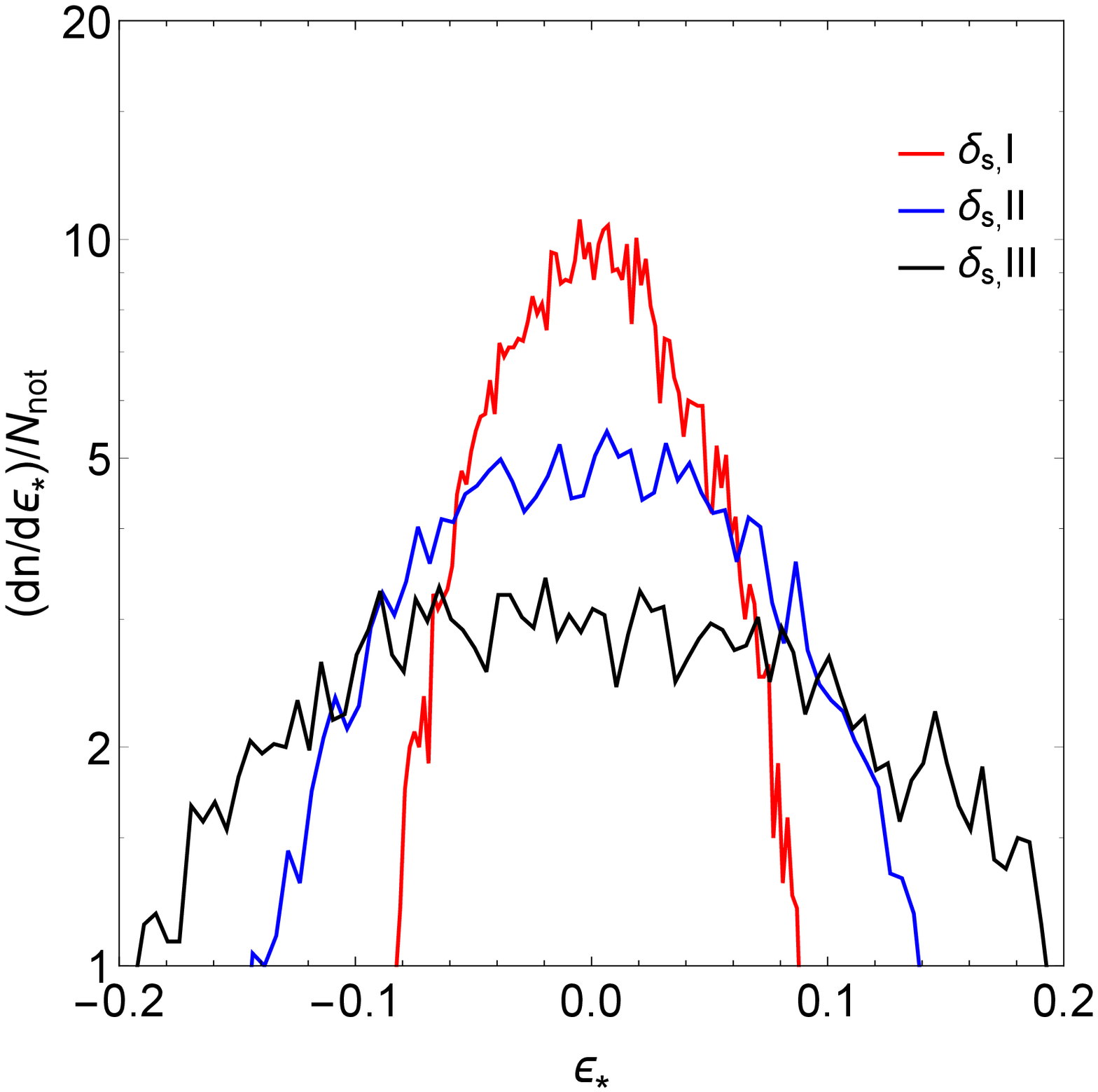}}
\caption{The probability density distributions of the measurement errors $\epsilon_{\ast}$ of three distortion parameters $\delta_{s,I}$, $\delta_{s,II}$, $\delta_{s,III}$. The shadow of a non-Kerr rotating compact object with quadrupole mass moment $q=-2,0.5,0.5,2$, and we set spin parameter $S=0.2M^{2}$.}
\label{sj02}
\end{figure}

\begin{figure}
\subfigure[$q=-2$]{ \includegraphics[width=4cm ]{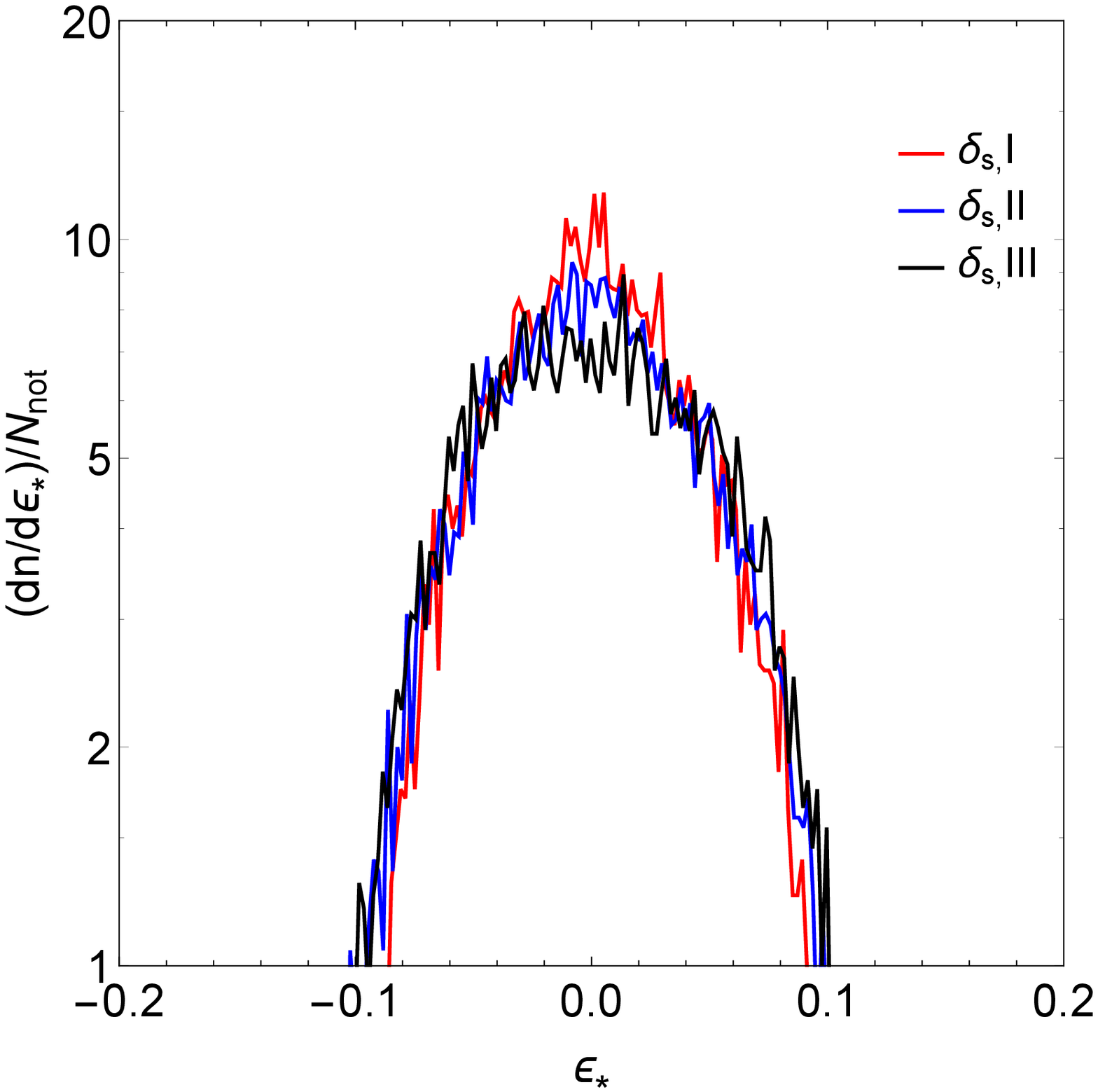}}\subfigure[$q=-0.5$]{ \includegraphics[width=4cm ]{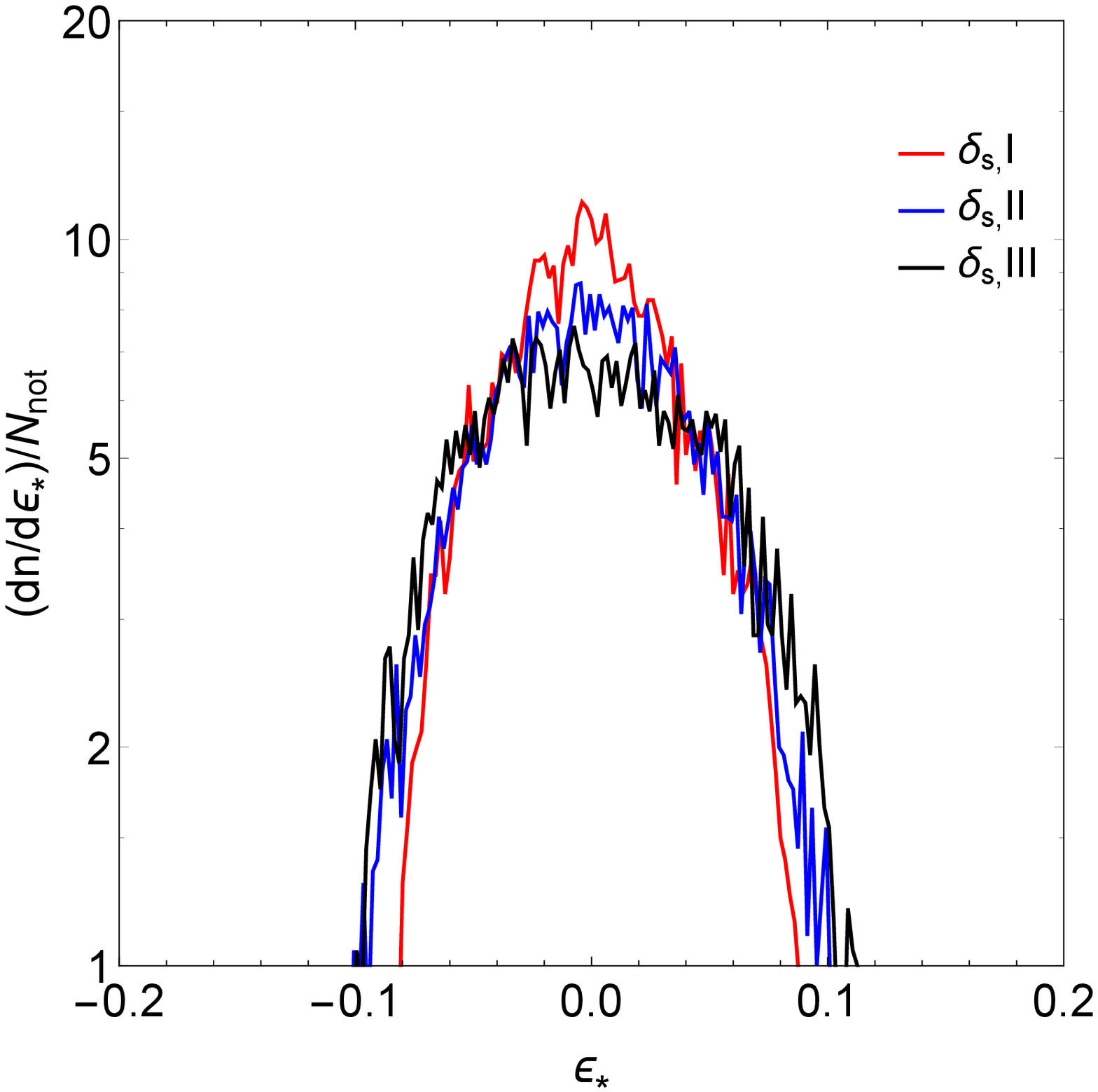}} \subfigure[$q=0.5$]{ \includegraphics[width=4cm ]{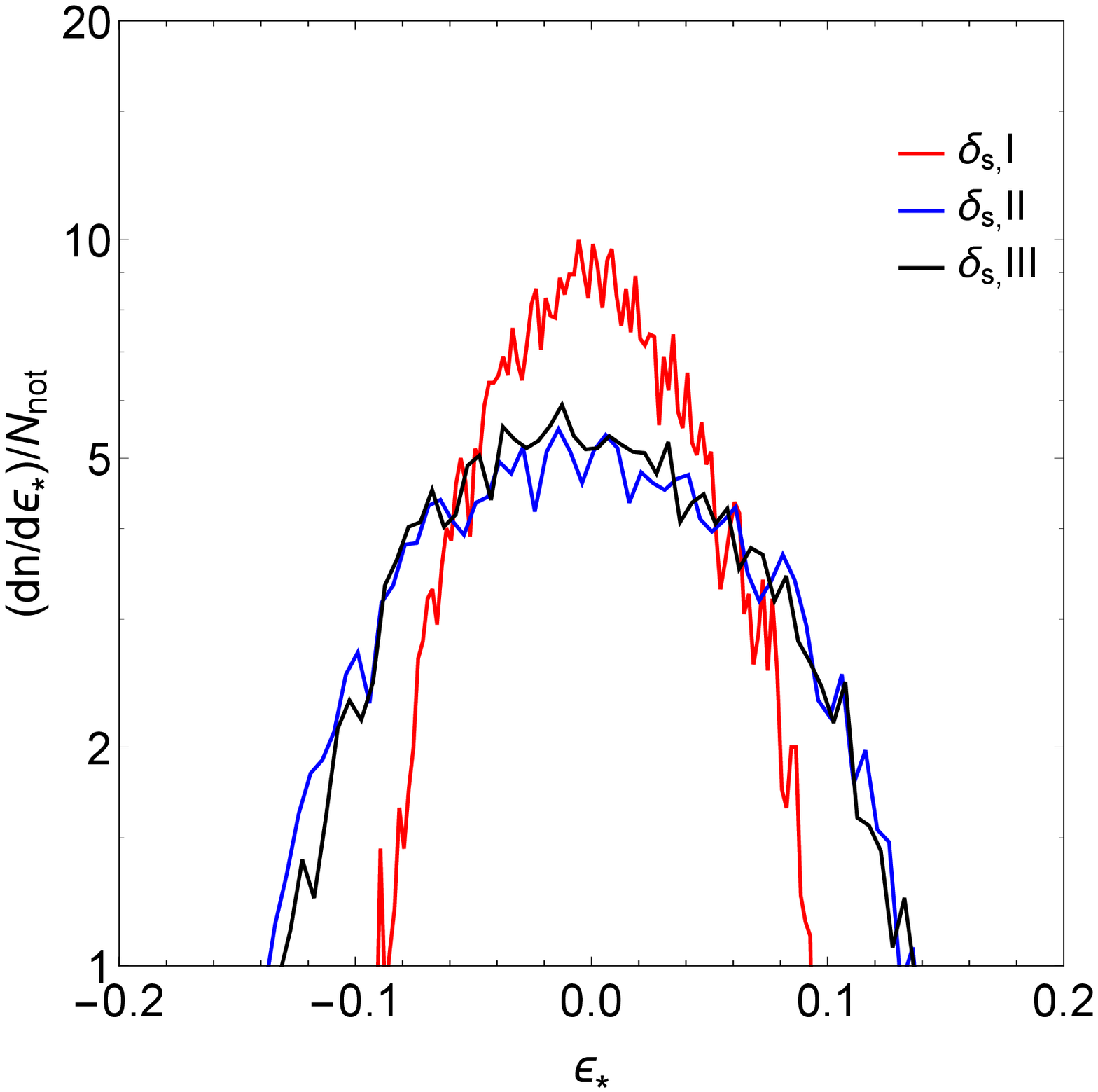}} \subfigure[$q=2$]{ \includegraphics[width=4cm ]{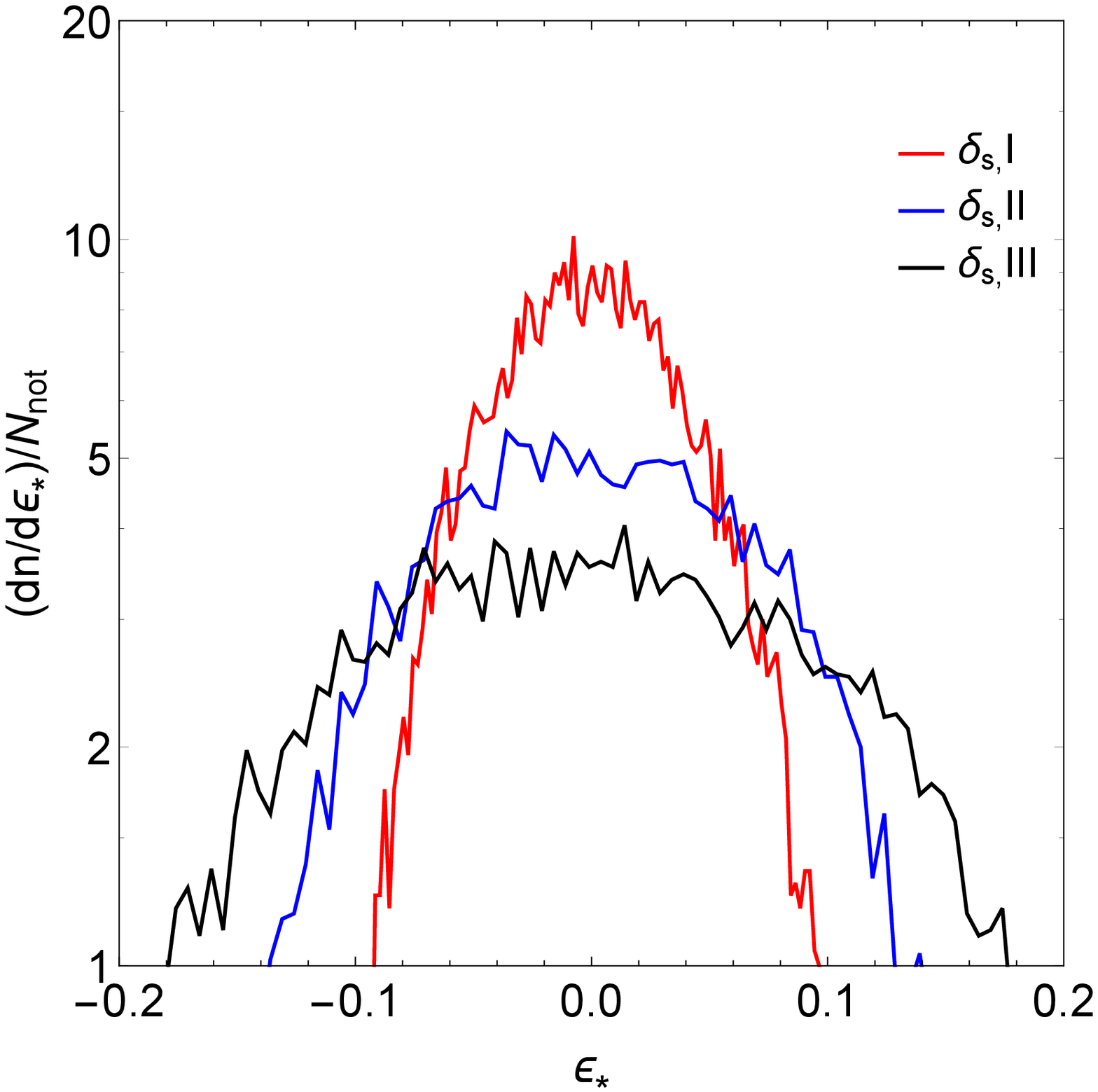}}
\caption{The probability density distributions of the measurement errors $\epsilon_{\ast}$ for three distortion parameter $\delta_{s,I},\delta_{s,II},\delta_{s,III}$. The shadow of a non-Kerr rotating compact object with quadrupole mass moment $q=-2,0.5,0.5,2$, and we set spin parameter $S=0.98M^{2}$.}
\label{sj098}
\end{figure}

\begin{figure}
\includegraphics[width=16cm ]{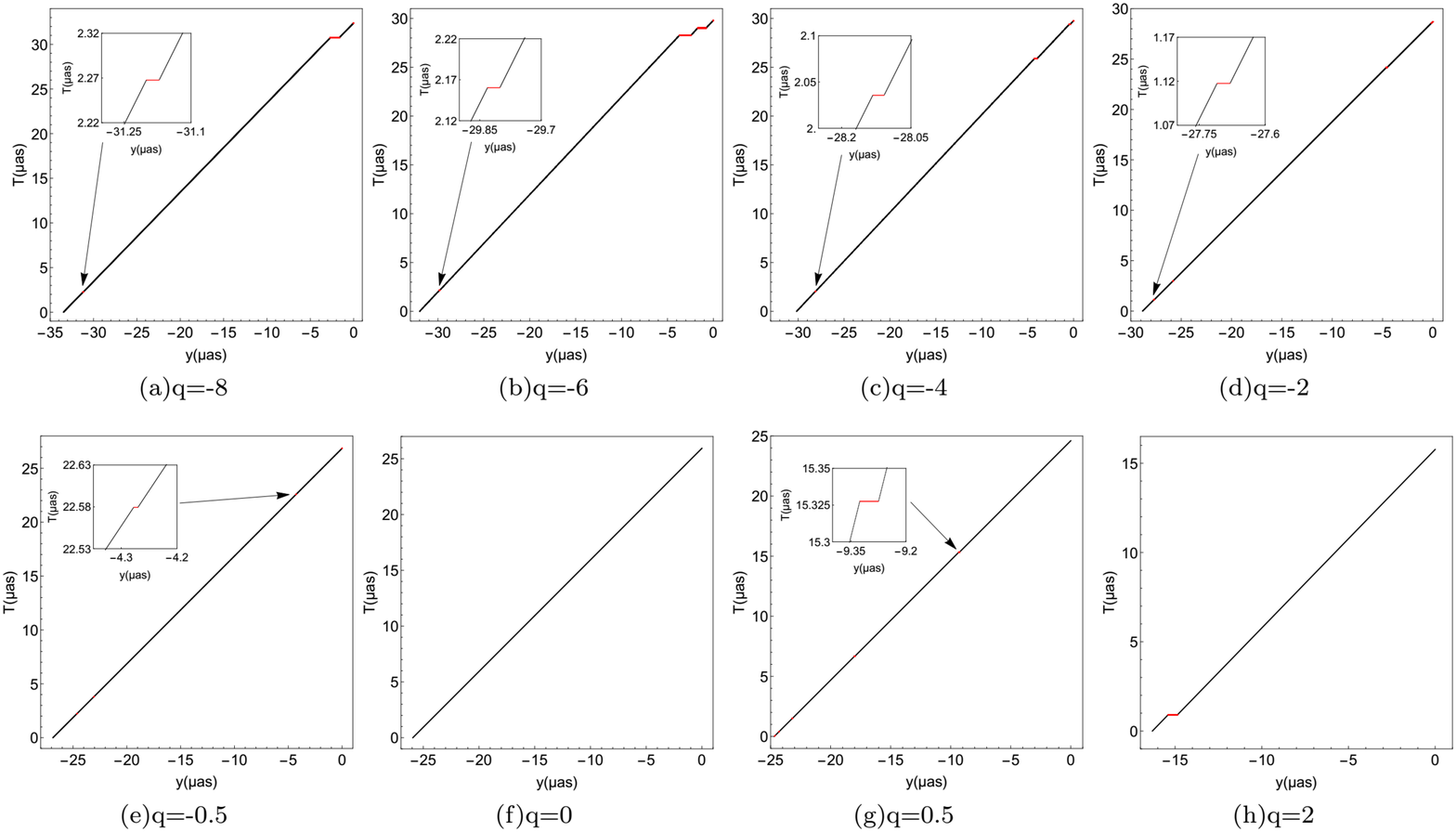}
\caption{The curves of $T(y)$ with quadrupole mass moment $q=-8,-6,-4,-2,-0.5,0,0.5,2$ for the fixed spin parameter $S=0.2M^{2}$ as we scan across the shadow along the line $x=0$.  Here we set observer's inclination angle $\theta_{obs}=90^{\circ}$.}
\label{pt02}
\end{figure}

\begin{figure}
\includegraphics[width=16cm ]{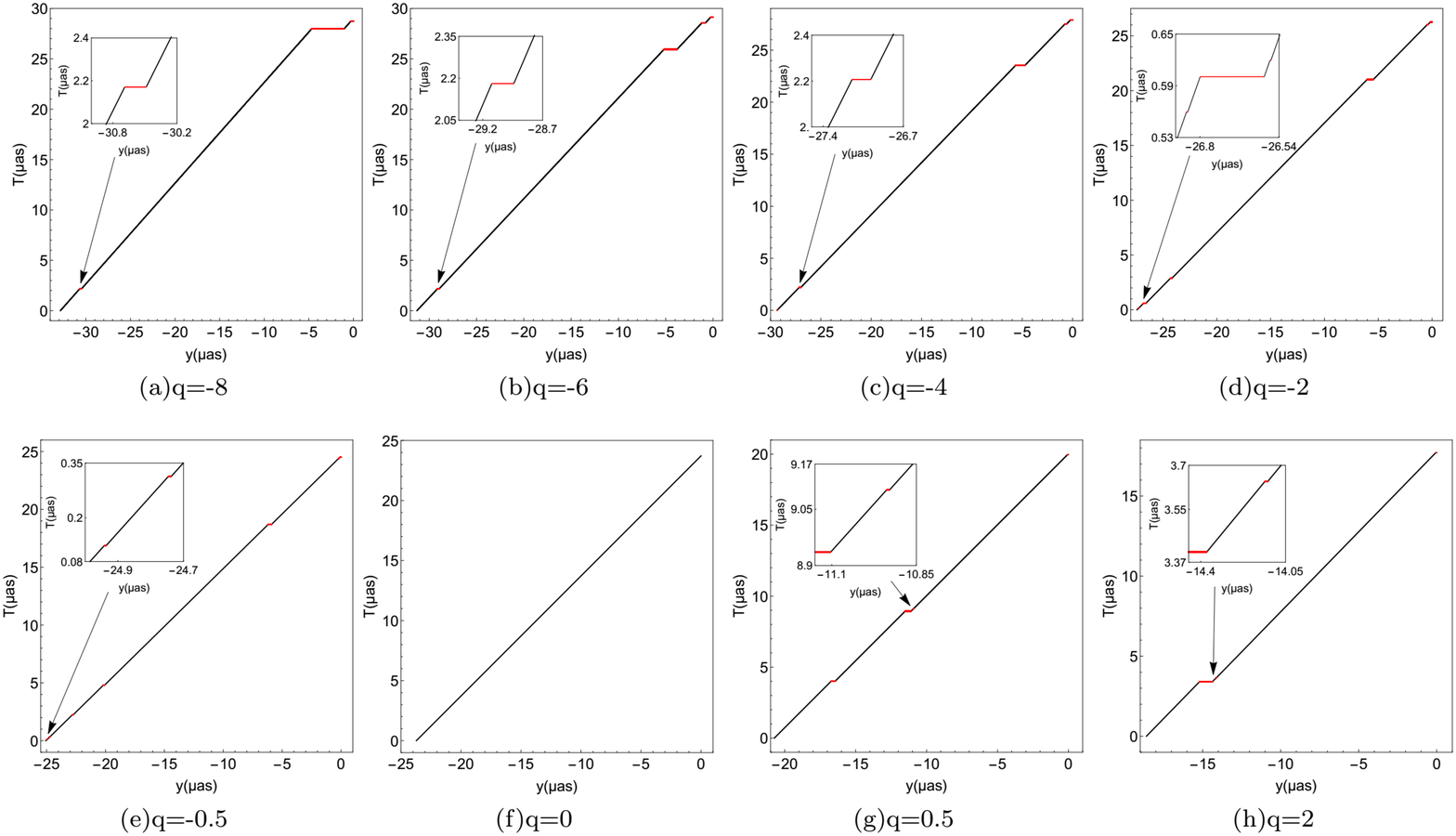}
\caption{The curves of $T(y)$ with quadrupole mass moment $q=-8,-6,-4,-2,-0.5,0,0.5,2$ for the fixed spin parameter $S=0.98M^{2}$ as we scan across the shadow along the line $x=0$.  Here we set observer's inclination angle $\theta_{obs}=90^{\circ}$.}
\label{pt098}
\end{figure}

\begin{figure}
\includegraphics[width=16cm ]{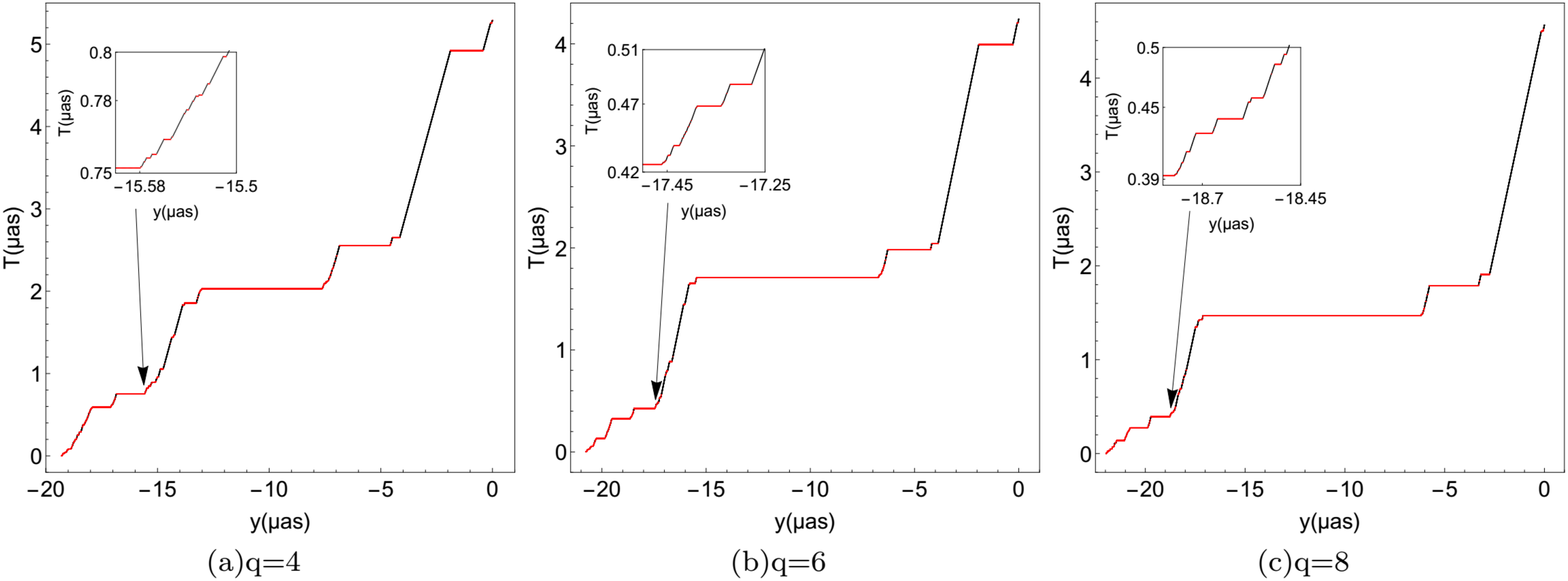}
\caption{The curves of $T(y)$ with quadrupole mass moment $q=4,6,8$ for the fixed spin parameter $S=0.2M^{2}$ as we scan across the shadow along the line $x=-4$.  Here we set observer's inclination angle $\theta_{obs}=90^{\circ}$}
\label{pt02h3}
\end{figure}

\begin{figure}
\includegraphics[width=16cm ]{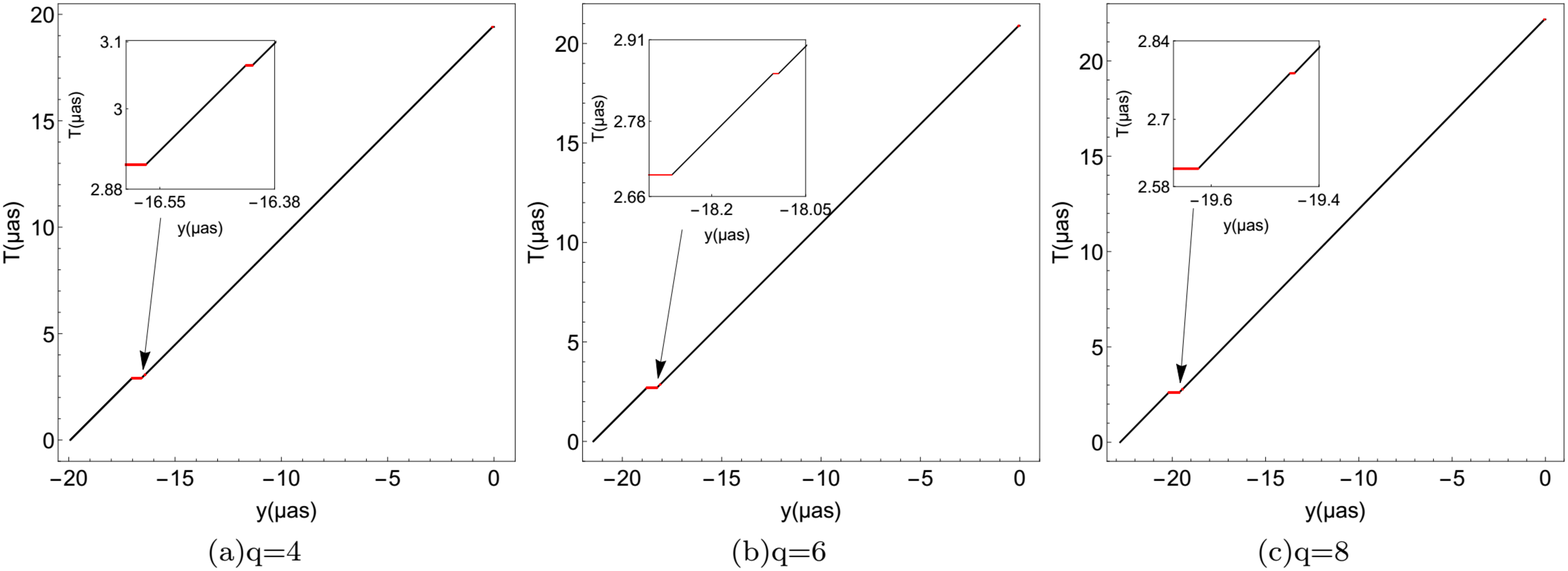}
\caption{The curves of $T(y)$ with quadrupole mass moment $q=-8,-6,-4,-2,-0.5,0,0.5,2$ for the fixed spin parameter $S=0.98M^{2}$ as we scan across the shadow along the line $x=-0.5$.  Here we set observer's inclination angle $\theta_{obs}=90^{\circ}$}
\label{pt098h3}
\end{figure}

The mass of the central object of our Galaxy is estimated
to be $4.4\times10^{6}M_{\odot}$ and its distance is around $8.5kpc$ \cite{sy49}. In tables II-III, we present the numerical values of observables $R_s$, $K_s$ and the magnitudes of three distortion parameters $\delta_{s,I}$, $\delta_{s,II}$, $\delta_{s,III}$ for the shadow of the supermassive black hole at the galactic center of Milky Way by applying the metric of the non-Kerr rotating black hole with quadrupole mass moment (\ref{MNdg1}) as the observer's inclination angle is $\theta_{obs}=90^{\circ}$ for different $q$ and $S$. Tables II-III indicate that these observables are functions of black hole parameters $q$ and $S$.
For fixed spin parameter $S$, one can find that $R_s$ decreases with $|q|$ for the negative quadrupole-deviation parameter, but it first decreases and then increases for the positive one. The oblateness parameter $K_{s}$ decreases with $|q|$ for the negative $q$. For the case with positive $q$, the change of $K_s$ with $q$ depends on the value of spin parameter $S$. Moreover, from tables II-III, one can find that $\delta_{s,I}$ first increases and then decreases with $|q|$ in the both cases $S=0.2$ and $S=0.98$. The changes of  $\delta_{s,II}$ and $\delta_{s,III}$ with $q$ become more complicated. For the negative $q$, the quantities $\delta_{s,II}$ and $\delta_{s,III}$ increase with $|q|$. For the positive $q$, $\delta_{s,II}$ and $\delta_{s,III}$ first increase and then decrease with $q$ in the case $S=0.2$, but they first decrease and then increase, and finally decrease with $q$ in the case $S=0.98$.
\begin{table}[ht]
\begin{center}
\begin{tabular}{|c|c|c|c|c|c|c|c|c|c|c|c|}
\hline \hline $q$ &$-8 $ &\; -6 &\;-4&\;-2&\;-0.5&\;0&\;0.5&\;2&\;4&\;6&\;8\\
\hline$x_{l}(\mu as)$ & -22.48& -21.97& -20.95&-20.44& -20.29& -23.91&-26.06&-14.46&-12.26&-11.24&-10.99\\
\hline
$y_{l}(\mu as)$ & $\pm$14.15& $\pm$13.03& $\pm$11.75&$\pm$9.96& $\pm$6.34& 0&0&0&0&0&0
 \\ \hline
$x_{r}(\mu as)$ & 25.14& 24.53& 24.02&23.50& 24.07& 28.10&29.64&32.60&34.75&36.79&38.94
 \\ \hline
$y_{r}(\mu as)$ & $\pm$14.31& $\pm$12.77& $\pm$11.50&$\pm$9.71& $\pm$4.85& 0&0&0&0&0&0
 \\ \hline
$x_{t},x_{b}(\mu as)$ & 0.05& 0.92& 1.28&1.64& 2.04& 2.55&2.76&13.34&15.84&16.61&16.86
 \\ \hline
$y_{t},-y_{b}(\mu as)$ & 33.42& 31.68& 30.15&28.97& 26.98& 26.06&24.83&21.46&21.46&22.48&23.35
 \\ \hline
$\Delta x(\mu as)$ & 47.62& 46.50& 44.97&43.94& 44.35& 52.02&55.70&47.06&47.26&48.03&49.92
 \\ \hline
$\Delta y(\mu as)$ & 66.84& 63.36& 60.29&57.94& 53.96& 52.12&49.67&42.92&42.92&44.97&46.70
\\ \hline $K_{s}$ & 0.71& 0.73& 0.75&0.76& 0.82& 0.998&1.12&1.096&1.095&1.068&1.069
\\ \hline $R_{s}(\mu as)$ & 34.80& 33.06& 31.36&30.13& 27.54& 26.07&24.91&21.59&21.63&22.61&23.39
\\ \hline $\delta_{s,I}$ & 0.1889& 0.2345& 0.2892&0.2106& 0.0510& 0.0021&0.1144&0.1377&0.1932&0.1869&0.1809
\\ \hline $\delta_{s,II}$ & 0.7661& 0.6100& 0.5697&0.5560& 0.3841& 0.0042&0.2457&0.3176&0.4476&0.3815&0.3658
\\ \hline $\delta_{s,III}$ & 0.6658& 0.5434& 0.5116&0.4968& 0.3500& 0.0045&0.2612&0.3870&0.8687&0.6712&0.5881\\
\hline\hline
\end{tabular}
\end{center}
\label{tab2} \caption{Numerical values of observables $R_s$ , $K_s$ and three distortion parameter $\delta_{s,I},\delta_{s,II},\delta_{s,III}$ for the shadow of the supermassive black hole at the galactic center of Milky Way with the metric of the non-Kerr rotating black hole with quadrupole mass moment
as observer's inclination angle is $\theta_{obs}=90^{\circ}$ for the fixed spin parameter $S=0.2M^{2}$.}
\end{table}
\begin{table}[ht]
\begin{center}
\begin{tabular}{|c|c|c|c|c|c|c|c|c|c|c|c|}
\hline \hline $q$ &$-8 $ &\; -6 &\;-4&\;-2&\;-0.5&\;0&\;0.5&\;2&\;4&\;6&\;8\\
\hline$x_{l}(\mu as)$ & -16.86& -15.84& -14.82&-13.18& -10.73& -11.75&-9.71&-9.71&-9.71&-9.81&-9.86\\
\hline
$y_{l}(\mu as)$ & $\pm$14.31& $\pm$13.29& $\pm$12.52&$\pm$11.24& $\pm$9.20& 0&0&0&0&0&0
 \\ \hline
$x_{r}(\mu as)$ & 29.79& 29.13& 29.13&29.23& 33.93& 35.26&36.28&38.53&40.37&42.41&43.94
 \\ \hline
$y_{r}(\mu as)$ & $\pm$14.21& $\pm$13.29& $\pm$ 11.24& $\pm$8.43& $\pm$0.10& 0&0&0&0&0&0
 \\ \hline
$x_{t},x_{b}(\mu as)$ & 7.26& 7.46& 7.66&7.92& 8.89& 10.07&13.29&18.29&20.95&21.46&21.77
 \\ \hline
$y_{t},-y_{b}(\mu as)$ & 33.72& 32.19& 30.15&28.97& 27.08& 26.06&24.78&23.30&22.99&24.02&24.73
 \\ \hline
$\Delta x(\mu as)$ & 46.65& 44.97& 43.94&42.41& 44.66& 47.01&45.99&48.24&50.08&52.22&53.81
 \\ \hline
$\Delta y(\mu as)$ & 67.45& 64.38& 60.29&57.94& 54.16& 52.12&49.56&46.60&45.99&48.03&49.46
\\ \hline $K_{s}$ & 0.69& 0.70& 0.73&0.73& 0.82& 0.90&0.93&1.04&1.089&1.087&1.088
\\ \hline $R_{s}(\mu as)$ & 36.50& 34.74& 31.90&30.35& 27.16& 26.08&24.85&23.53&23.32&24.24&24.88
\\ \hline $\delta_{s,I}$ &0.2094& 0.2730& 0.2207&0.2096& 0.1418&0.1073&0.1110&0.2155&0.2079&0.1945&0.1869
\\ \hline $\delta_{s,II}$ & 0.7087& 0.6950& 0.6318&0.6400& 0.4925& 0.2124&0.1324&0.1537&0.2120&0.2002&0.1849
\\ \hline $\delta_{s,III}$ & 0.6188& 0.6077& 0.5588&0.5649& 0.4605& 0.2108&0.1315&0.8386&0.7024&0.5728&0.4795\\
\hline\hline
\end{tabular}
\end{center}
\label{tab3}
\caption{Numerical values of observables $R_s$ , $K_s$ and three distortion parameters $\delta_{s,I}$, $\delta_{s,II}$, $\delta_{s,III}$ for the shadow of the supermassive black hole at the galactic center of Milky Way with the metric of the non-Kerr rotating black hole with quadrupole mass moment
as observer's inclination angle is $\theta_{obs}=90^{\circ}$ for the fixed spin parameter $S=0.98M^{2}$.}
\end{table}
Due to the existence of intrinsic uncertainties in astronomical observations, the white noisy could appear in the observational data for shadow's polar curve $R_{\psi}$. With the method proposed by A. Abdujabbarov \textit{et al} \cite{sha19s1}, we also analyse the observational random errors in the measurements of the shadow. In Figs.\ref{sj02}-\ref{sj098}, we present the probability density distributions of the measurement errors $\epsilon_{\ast}=\delta_{s}-\delta_{s,\ast}$ for three distortion parameters $\delta_{s,I}$, $\delta_{s,II}$, $\delta_{s,III}$ with different $q$ and $S$. Here $\delta_{s}$ is the exact distortion of the background metric (\ref{MNdg1}),  and  $\delta_{s,\ast}$ is given by $\delta_{s,I}$, $\delta_{s,II}$, $\delta_{s,III}$, respectively.
As in Ref. \cite{sha19s1}, all the distributions of the measurement errors centered on $\epsilon_{\ast}=0$, and the distortion paramete  $\delta_{s,I}$ has a smallest variations. Moreover, we find that the variance of the distortion parameters $\delta_{s,I}$, $\delta_{s,II}$, $\delta_{s,III}$ increase with quadrupole-deviation parameter $q$.

In Figs.\ref{pt02}-\ref{pt098h3}, we plot the curve of ``thickness" parameter $T(y)$ for different $q$ and $S$. For the case with $q=0$, there is only a single shadow for the black hole, we find that ``thickness" parameter $T$ increases linearly with the variable $y$ without any platform. For the cases with $q\neq0$, we find that there exist some platforms in the curve of ``thickness" parameter $T(y)$, which correspond to the bright region in the pattern of black hole shadow in which the light is not captured by black hole so that it can reach the observer. In Figs.\ref{pt02}-\ref{pt098}, we scan across the shadow along the line $x=0$ and find that the platforms appear in the cases with the quadrupole-deviation parameter $q=-8,-6,-4,-2,-0.5,0,0.5,2$. For the cases with $q=4,6,8$, we find that there exist also the platforms as we select the scan line $x=-4$ for $S=0.2M^{2}$ and the line $x=-0.5$ for $S=0.98M^{2}$. The width, positions and numbers of platforms depend on black hole parameters. The fractal structure in the platform set actually reflects  a self-similar feature of multiple shadows caused by chaotic lensing. These platform distribution could also provide a potential observable signal identified photon chaotic motion through black hole shadows.
Theoretically, comparing the theoretical values of these observables $R_s$, $K_s$, three distortion parameters $\delta_{s,I}$, $\delta_{s,II}$, $\delta_{s,III}$ and $T$ with those observation values obtained by astronomical instruments,  one can  extract the black hole information including the spin parameter $S$ and the quadrupole-deviation parameter $q$ from the black hole shadow. However, we note that resolutions of less than $0.1 \mu as$  are needed in order to extract useful information from observations of the shadow of the supermassive black hole at the galactic center of Milky Way, which is impossible for the current observational experiments, even for the most sensitive measurements, such as the Event Horizon Telescope \cite{charge2,charge3} and European Black Hole Cam \cite{charge4}.
We expect that it will reach in the future with the development of observation technology.

\section{Summary}

With the technique of backward ray-tracing,  we have studied numerically the shadows of a non-Kerr rotating compact object with an extra parameter related to quadrupole mass moment, which describes the deviation from Kerr black hole. Our result show that the shadows of a non-Kerr rotating compact object depend sharply on the quadrupole-deviation parameter $q$ and the rotation parameter $S$. For the case with negative $q$, the shadow of the non-Kerr rotating compact object with quadrupole mass moment becomes prolate and it is split into two disconnected  main shadows with eyebrows, which lie at symmetric positions above and below the equatorial plane.  With the increase of the deviation, the curvature of the boundary line near the equatorial plane increases for the main shadow. For the case with positive $q$, we find that the shadow of the non-Kerr rotating compact object with quadrupole mass moment becomes oblate and the main shadow are joined together in the equatorial plane. With the increase of the deviation, the shadow becomes oblate so that it turns into the shape of dumbbell for the larger $q$. In addition, there is a disorder region in the left of shadow which increases with the quadrupole-deviation parameter $q$. Interestingly, we also find that the Einstein ring is broken for the absolute value of $q$ more than a certain critical value $|q_c|$. The critical value $|q_c|$ decreases with the rotation parameter of non-Kerr rotating black hole.
We also discuss the dependence of the shadow on the observer's inclination angle $\theta_{obs}$. Especially, as $\theta_{obs}=0^{\circ}$, there are concentric bright rings appeared in the black disc. Our result show that the presence of quadrupole mass moment yields a series of interesting patterns for the shadow of a non-Kerr rotating compact object with quadrupole mass moment.

The model was applied to the supermassive black hole in the Galactic center. Our results show that for fixed spin parameter $S$, one can find that the observable $R_s$ decreases with $|q|$ for the negative quadrupole-deviation parameter, but it first decreases and then increases for the positive one. The oblateness parameter $K_{s}$ decreases with $|q|$ for the negative $q$. For the case with positive $q$, the change of $K_s$ with $q$ depends on the value of spin parameter $S$. We also study the effects of $q$ on the magnitudes of three distortion parameters $\delta_{s,I}$, $\delta_{s,II}$, $\delta_{s,III}$ for the shadow and analyse the probability density distributions of the measurement errors $\epsilon_{\ast}$ for these  distortion parameters with different $q$ and $S$. Finally, we note that there exist some platforms in the curve of ``thickness" parameter $T(y)$, which correspond to the regions in which the light is not captured by black hole and can reach the observer. The width, positions and numbers of these platforms depend on black hole parameters.

\section{\bf Acknowledgments}

We would like to thank the referee for useful comments.
This work was partially supported by the National Natural Science Foundation of China under
Grant No. 11875026, the Scientific Research
Fund of Hunan Provincial Education Department Grant
No. 17A124. J. Jing's work was partially supported by
the National Natural Science Foundation of China under
Grant No. 11475061, 11875025.

\end{document}